# Phase Curves of Nine Trojan Asteroids over a Wide Range of Phase Angles

Martha W. Schaefer
*Geology and Geophysics* and *Physics and Astronomy*
*Louisiana State University, Baton Rouge, LA  70803, USA*
*mws@lsu.edu*

Bradley E. Schaefer
*Physics and Astronomy*
*Louisiana State University, Baton Rouge, LA, USA*

David L. Rabinowitz
*Center for Astronomy and Astrophysics*
*Yale University, New Haven, CT, USA*

Suzanne W. Tourtellotte
*Department of Astronomy*
*Yale University, New Haven, CT, USA*

## Abstract

We have observed well-sampled phase curves for nine Trojan asteroids in B-, V-, and I-bands.  These were constructed from 778 magnitudes taken with the 1.3-m telescope on Cerro Tololo as operated by a service observer for the SMARTS consortium.  Over our typical phase range of 0.2-10°, we find our phase curves to be adequately described by a linear model, for slopes of 0.04-0.09 mag/° with average uncertainty less than 0.02 mag/°.  (The one exception, 51378 (2001 AT33), has a formally negative slope of -0.02±0.01 mag/°.)  These slopes are too steep for the opposition surge mechanism to be shadow hiding (SH), so we conclude that the dominant surge mechanism must be coherent backscattering (CB).  In a detailed comparison of surface properties (including surge slope, B-R color, and albedo), we find that the Trojans have surface properties similar to the P and C class asteroids prominent in the outer main belt, yet they have significantly different surge properties (at a confidence level of 99.90%).  This provides an imperfect argument against the traditional idea that the Trojans were formed around Jupiter's orbit.  We also find no overlap in Trojan properties with either the main belt asteroids or with the small icy bodies in the outer Solar System.  Importantly, we find that the Trojans are indistinguishable from other small bodies in the outer Solar System that have lost their surface ices (such as the gray Centaurs, gray Scattered Disk Objects, and dead comets).  Thus, we find strong support for the idea that the Trojans originally formed as icy bodies in the outer Solar System, were captured into their current orbits during the migration of the gas giant planets, and subsequently lost all their surface ices.

## Introduction

The Trojan asteroids share the orbit of Jupiter, orbiting the sun in two populations at the Lagrangian points of Jupiter's orbit, ~60° leading (L4) and ~60° trailing (L5) that body. The traditional idea was that these asteroids were captured rocky material from near the same orbit as Jupiter (Barucci et al., 2002). Alternatively, Morbidelli et al. (2005) suggested, based on dynamical reasons, that the Trojan asteroids were formed much farther out in the Solar System and were captured into co-orbital motion with Jupiter during the time when the giant planets themselves were migrating, just after Jupiter and Saturn crossed their mutual 1:2 resonance. If these bodies were formed at their present distance from the Sun, then their surfaces would always have been rocky. However, if they were formed much farther out in the Solar System and migrated inward, their surfaces would have originally been icy and would have de-iced since the migration. So we have a question: are Trojans captured asteroids or icy bodies captured during a migration of Jupiter?

If the Trojan asteroids were captured into their present orbits after their formation, then evidence of this should exist in their surface properties. Spectroscopy offers one method to study the dynamical evolution of these bodies. Recent work, for example by Bendjoya et al. (2004), Fornasier et al. (2004), and Roig et al. (2008) indicates that the Trojan asteroids are dominantly of the D taxonomic class, with a small fraction of the members of the P and C classes. Of the Trojans asteroids classified in the JPL Small-Body Database engine, 81% are of class D, 14% of class C, and 14% of class P (there is some overlap). Also from this database, the classified outer main belt asteroids have 20% in the D class, 34% in the C class, and 28% in the P class.

Another way to distinguish the origin of the Trojan asteroids is to compare their phase curves with those of regular asteroids and outer icy bodies. The phase curve of a body is a measurement of the reflected brightness of that body as a function of solar phase angle ($\alpha$, here presented in units of degrees), i.e., the angle from the Sun to the object to the Earth. In general, the body's brightness increases as it approaches opposition ($\alpha \sim 0°$) by as much as around half a magnitude. This brightening in the phase curve is called the *opposition surge*, and is the result of one or both of two mechanisms. The first mechanism is shadow-hiding, which is caused by the disappearance of shadows on a surface as the Sun gets behind the observer. The characteristic angular size of the opposition surge here is 6° or larger, and the brightening cannot be more than a factor of two in amplitude. The second mechanism, coherent backscattering, is caused by constructive interference of light rays multiply-scattered by the surface. The characteristic angular size of this surge is from a few tenths of a degree to perhaps 3°, and the amplitude is also restricted to be no more than a factor of two. In practice, surge amplitudes range from 0.03—0.5 mag, with coherent backscattering surges producing fairly sharp spikes in the phase curve, and the shadow-hiding surges leading to shallow slopes over observable ranges of phases (0—10° for Trojans). For general reviews on phase curves in our Solar System see Hapke (1993), Belskaya and Shevchenko (2000), Muinonen et al. (2002), Rabinowitz et al. (2007), and Schaefer et al. (2009a; b)

Phase curves provide one of the few independent means to compare the Trojans to possible parent populations. With this, we can compare Trojan phase curves with those of Centaurs, dead-comet candidates, and normal asteroids. If the traditional view of the formation of the Trojans is correct, then they should have phase curves comparable to those of normal asteroids from the outer main belt. If Trojans were originally icy bodies from the outer Solar System, then their phase curves should be more like those icy bodies that have suffered extensive volatile loss due to proximity to the Sun, such as dead comets and gray Centaurs. Schaefer et al., 2008b predict that the Trojans will have identical phase curves as the gray Centaurs and the dead-comet candidates.

Thus, the phase curves offer a possible means of distinguishing the origin of the Trojan asteroids. For the comparison, all available phase curves for the outer icy bodies have been collected in Schaefer et al. (2008a; 2009b), and phase curves for the asteroids are presented in Belskaya and Shevchenko (2000) and Muinonen et al. (2002). Prior work on phase curves of Trojan asteroids has been sparse. French (1987) reports on a slope for 1187 Anchises, but this is based on only four nights of data below 2°. What is needed are well-sampled phase curves for many Trojans.

To fill this need, and to test origin hypotheses for the Trojans, we have undertaken three campaigns to obtain well-sampled phase curves. Our photometry of 9 Trojans in the B, V, and I bands was all collected with the 1.3-m telescope on Cerro Tololo in the years 2006—2008. Our results are 9 well-sampled phase curves that we compare to asteroids, Centaurs, and dead comets.

## Data

For this project we used the 1.3 m telescope of the Small and Moderate Aperture Research Telescope System (SMARTS) consortium at Cerro Tololo. Observations were made by on-site operators at Cerro Tololo. Images were recorded with the optical channel (a Fairchild 2K × 2K CCD) of the ANDICAM (A Novel Dual-Imaging Camera), a permanently mounted, dual infrared/optical CCD camera. We binned the CCD in 2 × 2 mode to obtain $0.37''$ pixel$^{-1}$ and a $6.3' \times 6.3'$ field of view. Typical seeing is $1''$-$2''$. This telescope is queue-scheduled for shared use by all members of the SMARTS consortium; therefore we were able to obtain ~15 minutes of observing time per target for many nights around solar phase minimum out to near maximum phase. We usually observed two targets per night, often with a sequence of three exposures (B-V-I) per target. Telescope users share dome and sky flats, bias frames, and observations in B, V, and I of Landolt stars on all photometric nights. Our reduction procedure is described in detail elsewhere (Rabinowitz et al., 2006).

Given the time restrictions of the SMARTS queue schedule, we were limited to observing roughly four targets in any one campaign. We made a total of three campaigns, corresponding to the opposition of the L4 cluster in 2006, the L4 cluster in 2007, and the L5 cluster in 2008. Eurysaces, in the L4 cluster, was observed during two campaigns. We selected our targets as being those which have maximum apparent magnitudes brighter than about 18 and which passed through phases of less than 0.2° (although observing conditions prevented us from obtaining data at below 0.2° in a few cases). In all, we obtained good phase curves for 9 Trojan asteroids: 588 Achilles, 1208 Troilus, 4348 Poulydamas, 6998 Tithonus, 8317 Eurysaces, 12126 (1999 RM11), 13323 (1998 SQ), 24506 (2001 BS15), and 51378 (2001 AT33).

The basic properties of our 9 targets are summarized in Table 1. The number and name (or designation) of each asteroid is given, followed by which group it belongs to (the L4 group, generally named after the Greeks, or the L5 group, generally named after the Trojans). The absolute magnitude (H) and orbital inclination (i) follow, as taken from the NASA Horizons web interface (http://ssd.jpl.nasa.gov/horizons.cgi). The last three columns contain the range of dates over which observations were made, the solar phase range of the observations, and the total number of observations (including all three bands as separate observations).

All of our observed magnitudes are presented in Table 2. Due to the length of this tabulation, the table is only presented in electronic form, while a stub of this table (the first 10 lines) is presented in the paper copy, so that readers can see the format and contents of the table.

The first three columns are light-time corrected observation times (in JD minus 2450000) of the observations in the B, V, and I bands. The raw magnitudes and their errors in the three bands are the next six columns. The reduced magnitudes (the apparent magnitudes with the modulation due to the changing distance to the Earth and Sun removed) are listed in the next three columns. Thus, the reduced magnitude is $Bred = B - 5 \log_{10}(r\Delta)$, where r and $\Delta$ are the distances of the target to the Sun and Earth, respectively (in AU). The phase angle (Sun-asteroid-Earth) is in the final column.

The quoted uncertainties in Table 2 are for the measurement errors alone, and these are largely a product of the photon statistics in our images. However, there are additional systematic errors that have to be accounted for to get realistic error estimates. Part of this is an irreducible measurement error of 0.015 mag caused by variations in the CCD chip and recognizable by the scatter in the calibration of standard stars (Landolt, 1992; 2009). Therefore, in our analysis below, we always use the quoted statistical errors added in quadrature with the 0.015 mag systematic uncertainty. We are always using the same telescope and the same filters and the same CCD chip, so this provides us with a remarkable stability over both the short term and the long term. Standard stars (Landolt, 1992) are observed within a few hours (sometimes within a few minutes) of our Trojan fields, and these measures (using large photometric apertures to get nearly all the stars' flux and make the measures insensitive to changes in seeing, focus, or imperfections in centering) are used to calibrate the on-chip comparison stars for use in differential photometry (using small photometric apertures) with respect to the Trojan. We use multiple comparison stars that are substantially brighter than the Trojan, so the added statistical uncertainty from this process is always much smaller than the quoted error bars. The largest uncertainty in this process results from the correction for airmass in deriving the magnitudes of the comparison stars. The Trojan fields are almost always within a third of an airmass of the Landolt fields. The one-sigma scatter in the fluctuations of the extinction coefficients is 0.04, 0.03, and 0.02 mag per airmass superposed on seasonal variations of total amplitude of 0.03, 0.03, and 0.03 mag per airmass in the B-band, V-band, and I-band respectively (Gutierrez-Moreno et al., 1986; Landolt, personal communication; Rufener, 1986). With this, we expect typical errors in the calibration of the comparison stars caused by uncertainties in the extinction to be around 0.01 mag. But we must always be aware that extinction can change rapidly and some nights can have anomalously high extinction, and this can lead to an occasional magnitude with a substantially larger systematic error. Such errors are essentially impossible to know in practice, so we cannot add them in systematically, nor can we list them individually for each magnitude. While expected to generally be relatively small, this extinction calibration systematic error will be subsumed into the likely larger uncertainty associated with the rotation of the Trojan.

Therefore, we have added an arbitrary value for the systematic error ($\sigma_{sys}$) in quadrature with our measurement error to produce an estimate of the total error bar. The value of the systematic error for each asteroid has been chosen such that our model fits (see next section) produce a reduced $\chi^2$ of near unity. As such, the fact that our model fits have reduced $\chi^2$ near unity is by construction. Our derived $\sigma_{sys}$ values for each of the nine targets are presented in the third column of Table 3. For some of our asteroids (for example, 588 Achilles), the $\sigma_{sys}$ value is sufficiently small that any rotational modulation effects must be negligible. However, one of our asteroids (51378 (2001 AT33)) has an unexpectedly large $\sigma_{sys}$, which indicates substantial rotational modulation, for which we have not been able to determine a period. Nonetheless, we show below ("Rotational Modulation") that any rotational modulation is unlikely to significantly alter the slope we derive for the phase curve.

The reduced magnitudes ($B_{red}$, $V_{red}$, and $I_{red}$) in B, V, and I taken sequentially during a single night were averaged together to produce a single magnitude $m_{avg}$. The combination of the three filters is justified by the lack of any significant difference between the slopes. This color neutrality of the surges has a good precedent that all other known surges from small/gray objects are closely neutral (Schaefer et al., 2009b). Also, the excellent phase curve for Achilles shows a difference in slope of only 0.002 mag/° from V-band to R-band (Shevchenko et al., 2009). The JD of these observations were similarly averaged. If BVI sequences were taken more than once during a night, two magnitudes ($m_{avg}$) are reported. In cases where only a partial sequence was observed, the color of the object is assumed constant with time, and the mean B-V, V-I, or B-I (depending on which values were available) was used to calculate the missing value, which was then averaged as before. This procedure was performed in about a third of the BVI triplet sequences (almost all 8317 Eurysaces and 13323 (1998 SQ).

## Analysis

Phase curves are plots of the reduced magnitudes versus the phase angle. The error bars on our individual magnitudes are taken to be the total uncertainties, i.e., including the systematic error bars as described above. Our derived phase curves are plotted for all nine objects in Figures 1—9.

Our phase curves are well-sampled in phase and cover essentially all the observable range of phase angles. These phase curves show the usual opposition surge with a brightening towards low phase angle. The only exception is for 51378 (2001 AT33), which apparently is dimming as zero phase is approached.

Many asteroids display rotational modulation with significant amplitudes. These variations are uncorrelated with the phase angle and hence, if uncorrected, will introduce perhaps substantial scatter in any phase curve. We expect that uncorrected rotation effects will indeed be the dominant source of scatter in our phase curves. Of these nine Trojans, only 588 Achilles (Shevchenko et al., 2009) and 1208 Troilus (Molnar et al., 2008) now have known rotational periods, and we have not been able to determine any further confident rotational modulation in our data. We have performed Fourier transform analyses of all our light curves and all targets always have some highest peak in the Fourier transform. These highest peaks might be the real rotational period (or an alias of the real period), but we cannot prove that this is the case for any of the targets. Nevertheless, we have decided to remove the putative rotational modulation for the highest Fourier peak for all targets. However, this removal does not have a significant influence on the calculated phase slopes. Figures 1—9 show the phase slopes before the removal of the putative rotational modulation.

We fitted all of the phase curves to two models: a simple empirical linear fit, and a fit to the Hapke model (Hapke, 2002) for opposition surges. For empirical models, we chose the linear fit because it is the simplest possible model for the surge and this provides a fully adequate representation of the data. We also chose to fit the Hapke equations because this is a physics-based model. A difficulty with this model is that the resultant fit parameters are somewhat degenerate, for example between the contribution from shadow hiding (SH) and coherent backscattering (CB). We have chosen to not fit the H-G formalism because there is no physics in it, while as a purely empirical functional form, there is no reason to go to this complexity when a simple line fits well.

The linear fits to the phase curves were made by minimizing the $\chi^2$ comparison between the model and the data. The model equation is $m_{avg} = m_0 + S\alpha$, where $\alpha$ is the phase angle (in °), S is the slope of the phase curve (in mag/°), $m_0$ is the absolute magnitude of the body at zero phase angle, and $m_{avg}$ is our average reduced magnitude. The best fit values for all the data are presented in Table 3. We have also performed fits for just the data that have $\alpha \leq 2°$ and for $\alpha > 5°$, with the resulting slopes $S_{\leq 2}$ and $S_{>5}$ being tabulated in Table 3.

The columns in Table 3 are as follows: name of asteroid; number of observations ($N_{obs}$, considered to be the number of $m_{avg}$ for each asteroid); $\sigma_{sys}$ (described above); and the linear fit parameters: average phase slope (S) in mag/°, absolute magnitude at zero phase angle ($m_0$) in mag, the $\chi^2$ of the linear fit, and the other slopes $S_{\leq 2}$ and $S_{>5}$ in mag/°; the parameters of the Hapke fit are described below. The $\chi^2$ values for the linear and Hapke fits are such that the smaller of the two reduced $\chi^2$ values is unity by construction, as described above.

The average slope, S, is 0.06 mag/°, with a range of 0.04—0.09 mag/°, and with a surprisingly low RMS scatter of 0.02 mag/°. (For the statistics in this paragraph, we will ignore 51378 (2001 AT33), which has an anomalously negative slope.) The average slope at phase less than or equal to 2° is not significantly different than at phase greater than 5° (both 0.05 mag/°). This implies that our slopes do not greatly deviate from linearity. (Indeed, this will be seen later as the best-fit Hapke models do not have significantly better $\chi^2$ values than do the best-fit linear models.) While the linear models are good approximations to the observations, they do not reflect the known physics of the surge mechanisms.

Phase curve data were also fit to a Hapke model (Hapke, 2002) as follows. For our case of small phase angles, the observed magnitude is

$$m = m_* - 2.5 \times \log_{10}\left[p(1 + B_{S0}B_S) + M\right] - 2.5 \times \log_{10}\left[1 + B_{C0}B_C\right], \text{where}$$

$$B_S = (1 + \left[\tan(\alpha/2)/h_S\right])^{-1},$$

$$B_C = 0.5(1 + \left[(1 - e^{-Z})/Z\right])/(1 + Z)^2,$$

$$Z = \left[\tan(\alpha/2)\right]/h_c.$$

Here, $m_*$ is a magnitude that has absorbed various constants, p is the single particle scattering function evaluated for $\alpha\sim0°$, and M is the multiple scattering contribution. The contribution to the opposition surge from shadow hiding is given by $B_S$ with an amplitude of $B_{S0}$ and a width of $h_S$. The contribution to the surge from coherent backscatter is given by $B_C$ with an amplitude of $B_{C0}$ and a width of $h_C$. There are limits on the amplitudes from the physics of both mechanisms, such that $B_{S0} \leq 1$ and also $B_{C0} \leq 1$.

Through $\chi^2$ minimization, we tested several models for the opposition surge of these bodies: shadow-hiding only, coherent backscattering only, and coherent backscattering with maximal shadow-hiding. For example, for our best-observed target (8317 Eurysaces) we fitted a variety of surge models. For shadow-hiding alone, our best fit with a restriction that $h_S \geq 6°$ results in a $\chi^2 = 346$. This $\chi^2$ can be substantially improved (to $\chi^2 = 245$) by lowering $h_S$ to 1.2°, but this is unphysical, so we do not accept this fit. For coherent backscattering alone, we find a much better fit (with $\chi^2 = 106$) for $h_C = 4.6°$ and the $B_{C0}$ value pegged at 1.0. When $B_{C0}$ is lowered below this value, the $\chi^2$ is always increased. The comparison of these results (SH alone vs CB alone) for 8317 Eurysaces and the others observed here demonstrates that shadow-hiding does not dominate the opposition surge of the Trojans.

When shadow-hiding is added to the coherent backscattering the best fits are little influenced by varying the SH parameters. For Eurysaces, our best fit SH+CB model produces a $\chi^2 = 98$, for $h_C = 7.3°$, $B_{C0}$ pegged at 1.0, and $h_S$ pegged at 6°, and $B_{S0}$ pegged at 1.0. The reason for this is that the SH contribution is relatively flat, and at best can contribute only a small tilt to the modeled phase curve. Because our observed phase curves are largely linear, with small curvature, we cannot evaluate the SH contribution because the model fit parameters become degenerate. However, the overall contribution from SH must be small compared to the CB contribution, and can be neglected. For simplicity, we will only be reporting on the best fit model for CB alone. Our quoted error bars will only account for the uncertainties within the CB-only model, and these uncertainties would rise somewhat if we allowed a range of models including SH.

Our fits to the data with the Hapke model are summarized in Table 3. The final four columns in Table 3 present the best-fit Hapke parameters for the absolute magnitude at zero phase angle ($m_0$) in mag, coherent backscattering width ($h_C$) in degrees, coherent backscattering amplitude ($B_{C0}$), and the $\chi^2$ of the fit. We find that in all cases the amplitude of the coherent backscattering is always pegged at its maximum value ($B_{C0}$ = 1). The average width for coherent backscattering is $h_C$ = 7.5°, with a range from 4.5° to 10.3°. The object 51378 (2001 AT33) has a negative slope and no realistic Hapke fit.

We do not understand why the phase curve of 51378 (2001 AT33) has a slightly negative slope. It might be that there are additional phenomena that counteract the opposition surge, but we do not know of any such mechanism. We have no reason to believe that there are systematic errors, because the negative slope would require correlated errors in magnitudes determined independently on many nights. It is not impossible that rotational modulation could have this effect, but we have found no evidence for such. We know of no precedent for a negative phase slope among any asteroids or icy bodies. Perhaps the answer is just that the real phase curve is nearly flat, and ordinary random measurement errors make it appear to have a slightly negative slope. There is precedence for such a flat phase curve for one other small gray body -- the centaur 2002 GZ232 (Schaefer et al., 2009b). Based on this, we can expect about 1 in 10 of the small grey centaurs to have completely flat phase curves, which is consistent with what we are seeing for the Trojans.

## Rotational Modulation

All asteroids rotate, typically with periods from a few hours to a few days, so in principle they should all show some rotational modulation superposed on top of the phase curve. This modulation will be near zero for round and uniform bodies, while typical modulation amplitudes (peak to peak) are 0.1 mag or smaller, yet with some amplitudes getting up to a quarter or a third of a magnitude. Without correction, these rotational modulations will cause deviations from a smooth phase curve. Of the nine Trojan asteroids we observed, only 588 Achilles (Shevchenko et al., 2009) and 1208 Troilus (Molnar et al., 2008) now have known rotational periods, so we have to be concerned with the effects of uncorrected rotational modulation on our derived phase curves and slopes.

In general, uncorrected rotational modulation will not result in any significant changes in our results. The reasons are two fold: (1) we sample our phase curves with numerous observations spread widely in time and solar phase angle; and (2) rotational modulation is (almost) never correlated with the orbital phase. With this, the scatter in our plots will be somewhat larger, while the derived slopes and Hapke parameters will be nearly the same (although with somewhat larger error bars). We performed a quantitative analysis of this. In particular, we have measured the effects on the derived phase curve slope as the rotational period is varied from 0.1 to 10 days. We have adopted a sinusoidal modulation for this analysis. The analysis has been performed for the exact times and phases as observed for each of the Trojans. For our best-sampled Trojan (Eurysaces), for a peak-to-peak amplitude of 0.10 mag, the RMS scatter of the error in the phase curve slope is 0.0008 mag/°. All of the uncertainties related to rotational modulation will scale linearly with the amplitude, so a rotational modulation with an amplitude of 0.20 mag will have an added uncertainty of 0.0016 mag/°, and so on. For a much more poorly-sampled Trojan (Troilus), again with an amplitude of 0.20 mag (Molnar et al., 2008), the extra error in the slope will be 0.026 mag/°. In general, the extra uncertainty arising from uncorrected rotational modulation will be smaller than our reported uncertainties (see Table 3). The extra RMS scatter in our phase curves will be a factor of $2^{-1.5}=0.35$ times the amplitude. If all of the scatter in our phase curves above that introduced by the statistical errors in the magnitudes (i.e., $\sigma_{sys}$ as in Table 3) is caused by rotational modulation, then the amplitude should be near $\sigma_{sys}/0.35$. Thus, a number of our Trojans could well have amplitudes of near a quarter of a magnitude, leading to a larger than expected scatter in our phase curves. Nevertheless, for the results of this paper (essentially, the slopes of the phase curves), rotational modulation will not cause any significant changes.

There is one weak exception to this strong conclusion, and that is if the rotational period is close to one day (or half a day), so that the timing of our images (roughly at the same sidereal time each night) might end up being correlated with the phase. Let us illustrate this effect with a specific example where a Trojan has 70 days of observations before opposition all at the same sidereal time each night. If the rotational period is exactly one sidereal day, then our magnitudes would always be at the same part of the rotational modulation, and no change in the slope of the phase curve would result. But if the rotational period is 1/140 days shorter than the sidereal day, then the rotational modulation will change smoothly by exactly half a period over the observed time interval. In this case, if the rotational modulation happened to be near minimum at the start of the 70 day interval and hence near maximum at the end, then the resulting phase curve would have a systematic shift with the high phases being faint and the low phases being bright, resulting in an incorrect slope. In this extreme case, the error in the derived phase curve slope will be the amplitude divided by the range of the covered phase. For a typical amplitude of 0.10 mag and a phase range of 10°, the introduced error in the slope will be 0.01 mag/°. Such extreme cases are very unlikely because they require a fine-tuned period, a fine-tuned epoch of maximum light, and a large amplitude for the rotational modulation. The extreme cases are never reached because our observations have substantial scatter around some idealized identical sidereal time each night. Thus, under rare circumstances, a rotational modulation can produce a systematic change in the slope of the phase curve, but in practice, these effects must be small.

We can be quantitative on these effects for our Trojan phase curves. For the exact times and phases of our Eurysaces data, with a presumed rotational amplitude of 0.10 mag, the error in the slope of the phase curve will always be below 0.009 mag/° for all periods. As an illustration, for the exact times and phases of our Troilus data, if the rotational amplitude was 0.10 mag, the error in the slope would be less than 0.02 mag/° (our one-sigma uncertainty in the slope from Table 3) for periods outside the range of 0.452-0.454, 0.477-0.481, 0.908-0.926, and 0.962-0.981 days. In fact, Molnar et al. (2008) have found that the rotation period is 2.3405 days, so there is no correlation between orbital phase and rotational phase.

The phase curve for 51378 (2001 AT33) has a formally negative slope, and we had previously wondered whether this could be due to rotational modulation. For the exact times and phases of our data series, and for a rotational amplitude of 0.34 mag ($\sigma_{sys}$/0.35), we can get slope errors of 0.02 mag/° or larger for periods of 0.496-0.498, 0.501-0.502, 0.974-0.978, 0.984-0.995, 1.002-1.013, and 1.020-1.022 days. Again, there is less than a one percent probability that the true rotational period is in this range. If this possibility is realized, then the severe alias problems would prevent us from discovering the rotational period. The largest possible shift in the slope of the phase curve is 0.04 mag/°, so the rotationally corrected phase curve could have a positive slope up to 0.02 mag/°, which is still substantially flatter than all the other Trojans. Thus, 51378 remains an anomaly even with the unlikely possibility of systematic effects from rotational modulation.

Achilles can provide a test case to determine the importance of the rotational modulation. After our poster was presented at the 2008 DPS meeting in Cornell (Schaefer et al., 2008b), Shevchenko et al. presented an abstract for a well-observed rotational period and phase curve at the 2009 Lunar and Planetary Science Conference. The rotational modulation has a period of 7.306 hours, while showing a double-peaked light curve with the larger of the two peaks having an amplitude of 0.11 mag. For the 2007 opposition, the coverage is from 0.08°-9.72°, with the closely linear phase curves fitted with slopes of 0.045±0.001 and 0.043±0.001 mag/° in the V-band and R-band. This is identical with our result of a linear phase curve with a slope of 0.04±0.01 mag/° for phases 0.165°-8.202°. Shevchenko et al. (2009) used a massive observing campaign involving 13 people and four telescopes so as to get the slope accurate to 0.001 mag/°, whereas this degree of accuracy is not necessary for our calculations. So we view the excellent phase curve from Shevchenko et al. (2009) as good confirmation that our observations and methods are returning good phase curves (despite not having rotational corrections). Their results also provide further experience that the Trojans have phase curves that are closely linear and without significant variations with color.

The issue of correction for rotational modulation is an old one. Ideally, our community would like to have all phase curves corrected with a confident rotational light curve (for example, Shevchenko et al. (2009)), but this is generally not possible (for example, in the present work). Indeed, our field has an extensive literature of reported phase curves based on just a few magnitudes from within two or three small ranges of observed phases and no rotational corrections (e.g., Buratti et al., 1992; Degewij et al., 1980; Sheppard and Jewitt, 2002; Sheppard, 2007), for which any rotational effects are substantially worse than in this paper. The cost in time and money to get a rotational light curve (which requires many closely spaced nights of dedicated telescope time) in addition to a phase curve (which requires nights well spread throughout an observing season) is generally large. For most science questions (like the cause of the surge and the morphology of surges between classes of bodies), high accuracy (±0.001 mag/° versus ±0.01 mag/° or even ±0.03 mag/°) is not needed.

Our analysis has quantitatively demonstrated that our lack of rotational corrections only adds a small additional uncertainty in the slope of our observed phase curves, and these effects do not change any of our conclusions.

## Cause of the Opposition Surge

Two physical mechanisms have been proposed for the opposition surges, shadow hiding and coherent backscattering. To get physical information from observed phase curves, we must determine the cause of the surges. More correctly, as both causes must work at some level, we must determine which mechanism dominates over which phase ranges. Schaefer et al. (2009b) develop four criteria for distinguishing between the mechanisms based on observed properties. The first criterion is that CB must dominate if the surge is color dependent (i.e., the slope varies significantly with color), because shadows are essentially the same for all optical wavelengths. The Trojans have no significant color dependence, and this could arise from either SH or CB. The second criterion is that CB must dominate if the slope is significantly greater than 0.033 mag/°, with that being the maximal slope for SH due to its surge width being wider than 6° or so. This criterion is one that we can apply to our Trojan phase curves. The third criterion is that the exact shape of the phase curve will depend on the mechanism, as shown by the different shapes in equations in the Analysis section. Unfortunately, very high accuracy is required to make any useful distinctions, and so our Trojan phase curves cannot be used for this third criterion. The fourth criterion is that CB must dominate if the bodies albedo is larger than ~40%, with the reason being that high albedos will cause the shadows on the surface to be filled in and hence SH will have only small effect. The Trojans all have very low albedos, so this fourth criterion is not useful as either SH or CB could dominate for low albedo surfaces. In all, we are only left with one useful criterion, the second, for determining the dominant surge mechanism on the Trojans.

From Equation 1, the SH mechanism will have some steepest possible slope. Schaefer et al. (2009b) have derived the slope at zero phase to be $\approx 0.2(B_{S0}/h_s)$ mag/°. The physics constrains $B_{S0} \leq 1$. We have an additional constraint that $h_s \geq 6°$, either theoretically or empirically. The width of the SH surge will be given by the average size of the shadow caster divided by the length of the shadow cast. For normal geometry of a jumbled surface, the ratio cannot be too small as the porosity of surfaces is not high. Detailed theoretical models of packed particles return similar results (Hapke 1993). Empirically, for the cases where the SH and CB component can be distinguished, the typical value of $h_S$ ranges from 10°-40° and is never smaller than 6° (e.g., Buratti and Veverka (1985); Helfenstein et al. (1997); Stankevich et al. (1999); Verbiscer et al. (2005)). With these two constraints, the slope at zero phase can only be smaller than 0.033 mag/°. We are reporting on the phase curve slopes out to many degrees, and so the fitted slope over these extended ranges must be smaller than 0.033 mag/°. For slopes over phases less than 2°, shadow-hiding must have $S_{\leq 2}$ less than 0.026 mag/°, and for typical values of $B_{S0}$ the slope is usually only 0.01 mag/°. As such, we have a confident criterion for distinguishing between SH and CB; if the slope is significantly greater than 0.026 mag/°, then we can confidently identify CB as the dominant surge mechanism.

Of our nine Trojans, eight have slopes that are steeper than the maximal SH phase curve. Then, by our second criterion, the CB mechanism must be dominating. The one exception is 51378 (2001 AT33), for which the formal slope is slightly negative, for which neither the SH nor the CB models are consistent. In this exception, our second criterion cannot be applied, with some other effect (perhaps rotational modulation) dominating. So eight out of eight Trojans, for which our second criterion can be applied, have the CB mechanism dominating.

## Comparison with Other Groups

To determine the provenance of the Trojan asteroids, we compare them to other groups. The groups with which we can compare our sample are the main belt asteroids, the de-iced outer Solar System bodies, and the icy outer Solar System bodies. The main belt asteroid group objects we compare have surface taxonomy classes including D, P, C, M, S, and E. The de-iced outer Solar System objects (the 'Small/Gray' class of bodies) we include are the gray Centaurs, the gray SDOs, and the dead comet candidates. The icy outer Solar System bodies (the 'Small/Red' bodies) include Kuiper Belt objects (KBOs) and red Centaurs. No one expects that the Trojans will have surfaces like the icy outer Solar System bodies, but this group is compared here for completeness. We also include comparisons with the transition bodies between the Small/Gray and Small/Red classes, including active Centaurs and active comets.

To make these comparisons, we have collected a variety of properties for each class of bodies into Table 4. These include the surge slopes for phase angles $\leq 2°$ ($S_{\leq 2}$), the number of bodies for which this was measured ($N_{\leq 2}$), the surge slope $>5°$ ($S_{>5}$), the number of bodies for which this was measured ($N_{>5}$), the B-R color (corrected to the solar color index), and the albedo.

The Trojans have greatly different properties from the Small/Red bodies, with all measured properties being disjoint in range. Because the Small/Red bodies have icy surfaces, we only have the expected conclusion that the Trojans do not have icy surfaces. The Trojans have similar colors and albedos as the transition bodies, but only to the extent that the transition bodies are like the Small/Gray bodies. We have only one measured surge value for a transition body, and that is outside the range for Trojans but is within the range of the Small/Red bodies. The Trojans also have greatly different surges, colors, and albedos when compared to the M, S, and E asteroid types. We conclude that the Trojans do not share an origin with the main belt asteroids and are not simply 'captured' main belt asteroids.

The two competing hypotheses for the origin of the Trojans are that they formed at around the distance of Jupiter (the 'traditional' idea) or that they formed much farther out in the Solar System (Morbidelli et al., 2005). The Small/Gray bodies provide a good comparison sample for the far-out origin, and indeed, Schaefer et al. (2008a; 2009b) have made a specific prediction that the Trojans will have an identical distribution of surge properties as the Small/Grays. But we do not have any confident sample of bodies that formed about the orbital position of Jupiter. After all, the Trojans are the question at hand, and everything else has been scattered away. The best that we can do for a comparison sample is the D, P, and C class asteroids that predominate in the outer main belt. It is reasonable that these classes are the most similar to bodies formed around Jupiter's orbit. That is, perhaps the conditions in the outer main belt make for surfaces similar to those of bodies somewhat farther out. Indeed, most Trojans with spectral classification are placed into the D class (with the remainder being placed in the P and C classes). So, for testing the traditional idea that the Trojans formed around Jupiter's orbit, the best available group for comparison is the D, P, and C asteroids in the outer main belt, even though it is not quite the correct comparison.

How do the Trojans compare with the Small/Gray bodies in their surface properties? The two groups have indistinguishable colors and albedos (see Table 4). The Small/Gray bodies cannot be observed to high enough a phase angle, so we do not have measures of $S_{>5}$ for them. The surge slope at low phase angles ($S_{\leq2}$) has good overlap between the Small/Grays and the Trojans, but the Small/Grays appear to go to higher values than the Trojans. The range of $S_{\leq2}$ for all nine of our observed Trojans is -0.01 to 0.09 mag/°. The range of $S_{\leq2}$ for all nine small/gray bodies (8 gray Centaurs and 1 gray SDO) is 0.01-0.18 mag/°. These ranges are comparable, but three of the small/gray bodies lie above the range for the Trojans. The key science question is whether the two distributions of $S_{\leq2}$ are consistent with being drawn from the same parent population. A statistical test to answer this question is the Kolmogorov-Smirnov (K-S) test (Press et al., 1986). We have constructed the cumulative distributions in $S_{\leq2}$ for nine objects in each of the two samples, and we find the largest difference between the two distributions is 0.44. The K-S probability is then that two samples taken from the same parent population will have a difference ≤0.44 only 25% of the time. This is near a confidence level that corresponds to a Gaussian one-sigma result, and such differences are commonly expected. That is, the two distributions are fairly close and certainly not significantly different. Thus, we conclude that the Schaefer et al. (2008a; 2009b) prediction is confirmed, because the Trojans appear to have a similar distribution of surge properties as the other Small/Gray bodies in our Solar System.

What we are seeing is that all the surface properties of the Trojans are indistinguishable from those of the gray Centaurs, gray SDOs, and dead comet candidates. That is, all four dynamically distinct groups have the same surfaces. This could arise because the bodies are all genetically related, in that they all come from the same parent population and have all arrived at the same end–state (i.e., lost their surface ices) through differing evolutionary paths that leave them with different orbits. Thus, the bodies all formed in the outer Solar System (then appearing as Small/Red bodies), and were later perturbed into orbits that carried them close enough to the Sun for them to lose their surface ices and be transformed into Small/Gray bodies. Thus, for surface properties, it does not matter whether the ice was lost long ago after capture into a Trojan orbit long ago or whether the ice has been recently lost by a Centaur that is temporarily getting close enough to the Sun.

We can now test the traditional hypothesis that the Trojans formed around the orbit of Jupiter. Again, we only have the imperfect comparison with the D, P, and C class asteroids. The albedos of the Trojans and asteroids are indistinguishable. The B-R colors are similar, even though the Trojans appear to extend to slightly smaller values. But due to limited numbers of Trojans and D class asteroids, this difference in distribution is not significant. The distribution of $S_{>5}$ is similar for the two groups, but again the small numbers makes for a relatively weak test. For $S_{\leq 2}$, the ranges of the two groups are close to each other and with a small overlap. But the P and C class asteroids all have closely identical phase curves (Belskaya and Shevchenko, 2000) with slopes only at the top end of the Trojan range. We have used a K-S test to see if the measured $S_{\leq 2}$ are consistent with both samples being drawn from the same parent population. Our comparison is between our nine Trojans and the thirteen P and C class asteroids with phase curves in Belskaya and Shevchenko (2000). No D class asteroids in the main belt have measured phase curves. The two cumulative distributions have a maximum difference of 0.79, in that 79% of the Trojans have $S_{\leq 2}<0.065$ mag/° while 0% of the P and C asteroids have $S_{\leq 2}<0.065$ mag/°. The K-S probability is then that two samples taken from the same parent population will have a difference $\leq 0.79$ only 0.10% of the time. That is, at the thousand-to-one confidence level, the surge properties of the Trojans and the asteroids are different. This confidence level is substantially stronger than '3-sigma', so we take this as a strong result.

A plausible possibility is that the Trojans have a mixed population, composed of bodies originating from near the orbit of Jupiter and from the outer Solar System. We can use a K-S test to put limits on the fraction of Trojans from a population like the P and C asteroids. At the two-sigma confidence level (95%), we find that no more than one-third of the Trojans can come from the asteroid classes.

With this, we see that the Trojans do not have the same surface properties as the P and C asteroids. Thus, to the extent that the P and C asteroids are representative of the bodies that originally formed around Jupiter's orbit, we can exclude the Trojans as having come from that population. As such, we have evidence against the traditional view for the formation of the Trojans.

Nevertheless, we are intrigued by the close similarity in surface properties between the Small/Gray bodies and the P and C asteroids. In particular, if we put the P and C types together in one group, their phase slopes, B-R, and albedo distributions are indistinguishable from the small/gray group. We have performed a K-S test to any difference in the surge slope distribution between the 9 Small/Gray bodies and the 13 P and C asteroids. The maximum difference between the cumulative distributions is 0.36, with 44% of the Small/Gray bodies and 8% of the P&C asteroids having $S_{\leq 2}<0.075$ mag/°. The K-S probability is 62% of the two samples coming from the same parent population will have a maximum difference this large. This is saying that the two distributions are indistinguishable. (Note, that for surge slopes, the Trojans and P and C asteroids can both be indistinguishable from the Small/Grays and yet be significantly different from each other.) Viewed on the basis of the surface properties we have examined, the P and C asteroids look exactly like the Small/Gray bodies of the outer Solar System.

Very recently, Levison et al. (2009) have suggested that most of D and P types in the main belt formed in the outer Solar System and were captured, like the Trojans, during an early phase of planetary migration. Our observations that P- and D-type asteroids, Trojans, and small gray bodies from the outer Solar System not only have similar colors and albedos but also similarly flat phase curves supports this conclusion. However, the somewhat different surge slopes between the Trojans and the P and C asteroids suggests that there might be some differences in history or original composition. Within this picture, the large class of Small/Gray bodies (that is, small bodies formed in the outer Solar System that have lost all their surface ices) consists of gray Centaurs, gray SDOs, dead comets, Trojans, D-type asteroids, P-type asteroids, and C-type asteroids.

In summary, we have nine well-measured phase curves, the first for Trojan asteroids. Our phase curves are nearly linear with median slopes of 0.05 mag/°. These slopes are too high for the shadow-hiding mechanism of opposition surge, so the coherent-backscattering mechanism must dominate. In a comparison with the Small/Gray bodies (i.e., objects formed in the outer Solar System that have lost their surface ices), we find that the Trojans have a similar distribution of surface properties. This is a fulfillment of a prediction that the Trojan surges would be indistinguishable from the Small/Gray surges. Indeed, we find that the Trojans are indistinguishable from the Small/Gray bodies in all surface properties. In a comparison with all the other classes of bodies (the Small/Red icy bodies, transition bodies, and five main belt asteroid classes), we find that the surge properties are significantly different. This result is evidence against the traditional idea that Trojans formed around the orbit of Jupiter. Overall, our results are in good agreement with the theory that the Trojans are simply de-iced bodies from the outer Solar System (Schaefer et al., 2008b; 2009b) and were captured into their current orbits during the migration of the giant planets (Morbidelli et al., 2005).

## Conclusions

Trojan asteroids can serve as an indicator for the early history of our Solar System, with the Nice model of Morbidelli et al. (2005) claiming that the Trojans were originally formed in the outer Solar System, in contrast to the traditional view that they formed near the orbit of Jupiter. If Trojans can be demonstrated to come from the outer Solar System, then this would lend good support to the details of the migrations of the gas giant planets. One way to test the Nice model is to look at the surface properties of the Trojans in comparison with de-iced bodies from the outer Solar System (including the gray Centaurs and dead comets) and with asteroids from the outer belt (like the P and C classes).  To this end, we are reporting on a program to get well-sampled phase curves of Trojans.

Prior to our work, there was essentially no useful measure of the phase curve of any Trojan asteroid.  (After our work, Shevchenko et al. (2009) report a remarkably accurate phase curve for one Trojan, which is in close agreement with our phase curve.)  Now, with our work, we have nine good phase curves for comparison with other classes of bodies.

We find that the nine Trojan phase curves are consistent with a simple linear model from phases over the range of roughly 0.2°-10°.  The slopes for eight of them are from 0.04-0.09 mag/°.  (The ninth Trojan has an unprecedented negative slope of -0.02±0.01 mag/°, for which we have no sure explanation.)  The average uncertainty in these slopes is 0.02 mag/°, although there could be some extra systematic uncertainty at this level or smaller arising from uncorrected rotational modulations.  With these slopes, the coherent backscattering mechanism must be dominating over the shadow hiding mechanism.

We compared the Trojan surface properties (surge slope, albedo, and B-R color) with those of de-iced outer Solar System bodies as well as those of outer main belt asteroids.  For the albedo and color parameters, the Trojans are indistinguishable from either group.  For the surge slope, the Trojans are indistinguishable from the small/gray bodies, yet are significantly different from the P and C asteroids.  Thus, we find support for the idea that the Trojans were formed in the outer Solar System, captured into the Trojan orbit during migration of the giant planets, and then lost their surface ice due to proximity with the Sun.

## Acknowledgments

The National Aeronautics and Space Administration provided funds to support this research under grants NAG5-13533 and NAG5-13369.

**Table 1.  Summary of Observations**

| Number | Name | Group | H (mag) | i (°) | Dates observed | Phase range(°) | # obs. |
|--------|------|-------|---------|-------|----------------|----------------|--------|
| 588 | Achilles | Greek/L4 | 8.67 | 10.32 | 2007 July 12-Sep 30 | 0.165-8.202 | 119 |
| 1208 | Troilus | Trojan/L5 | 8.99 | 33.56 | 2008 May 5-May 24 | 0.199-3.247 | 35 |
| 4348 | Poulydamas | Trojan/L5 | 9.20 | 7.96 | 2008 Mar 22-May 13 | 0.207-6.821 | 26 |
| 6998 | Tithonus | Trojan/L5 | 11.30 | 1.73 | 2008 Apr 11-May 10 | 0.052-4.576 | 31 |
| 8317 | Eurysaces | Greek/L4 | 10.70 | 0.95 | 2006 Jun 13-2007 Oct 2 | 0.132-11.067 | 228 |
| 12126 | (1999 RM11) | Trojan/L5 | 10.10 | 2.04 | 2008 Mar 26-May 6 | 0.173-7.096 | 40 |
| 13323 | (1998 SQ) | Greek/L4 | 10.70 | 0.91 | 2007 Mar 21-Oct 2 | 0.349-10.33 | 91 |
| 24506 | (2001 BS15) | Greek/L4 | 10.50 | 11.82 | 2007 July 12-Sep 29 | 0.136-8.651 | 119 |
| 51378 | (2001 AT33) | Greek/L4 | 11.10 | 33.63 | 2007 July 11-Oct 2 | 0.214-10.28 | 114 |

**Table 2.**

**(a) 588 Achilles**

| JD (B) | JD (V) | JD (I) | B | $\sigma_B$ | V | $\sigma_V$ | I | $\sigma_I$ | $B_{red}$ | $V_{red}$ | $I_{red}$ | Phase |
|---|---|---|---|---|---|---|---|---|---|---|---|---|
| 4293.8445 | 4293.8465 | ... | 16.646 | 0.013 | 15.904 | 0.008 | ... | ... | 9.502 | 8.760 | ... | 5.584 |
| 4295.8272 | 4295.8292 | 4295.8311 | 16.657 | 0.007 | 15.889 | 0.005 | 14.946 | 0.010 | 9.521 | 8.753 | 7.810 | 5.266 |
| 4309.7259 | 4309.7279 | 4309.7298 | 16.431 | 0.043 | 15.620 | 0.023 | 14.672 | 0.031 | 9.345 | 8.534 | 7.586 | 2.826 |
| 4315.7208 | 4315.7228 | 4315.7247 | 16.420 | 0.013 | 15.610 | 0.011 | 14.684 | 0.016 | 9.347 | 8.537 | 7.611 | 1.691 |
| 4316.7279 | 4316.7299 | 4316.7318 | 16.336 | 0.018 | 15.560 | 0.005 | 14.705 | 0.009 | 9.265 | 8.489 | 7.634 | 1.497 |
| 4319.7276 | 4319.7296 | 4319.7315 | 16.326 | 0.006 | 15.549 | 0.004 | 14.657 | 0.004 | 9.260 | 8.483 | 7.591 | 0.916 |
| 4320.5518 | 4320.5538 | 4320.5557 | 16.253 | 0.007 | 15.551 | 0.005 | 14.664 | 0.010 | 9.188 | 8.486 | 7.599 | 0.756 |
| 4320.8455 | 4320.8475 | 4320.8494 | 16.268 | 0.008 | 15.543 | 0.007 | 14.677 | 0.009 | 9.203 | 8.478 | 7.612 | 0.699 |
| 4322.8415 | 4322.8435 | 4322.8454 | 16.228 | 0.009 | 15.548 | 0.011 | 14.590 | 0.006 | 9.165 | 8.485 | 7.527 | 0.315 |
| 4323.6699 | 4323.6721 | 4323.6742 | 16.292 | 0.048 | 15.548 | 0.041 | 14.634 | 0.048 | 9.230 | 8.486 | 7.572 | 0.165 |
| 4325.6666 | 4325.6686 | 4325.6705 | 16.308 | 0.008 | 15.541 | 0.007 | 14.624 | 0.009 | 9.247 | 8.480 | 7.563 | 0.264 |
| 4325.8304 | 4325.8323 | 4325.8343 | 16.261 | 0.010 | 15.547 | 0.010 | 14.608 | 0.010 | 9.200 | 8.486 | 7.547 | 0.295 |
| 4326.5248 | 4326.5270 | 4326.5291 | 16.296 | 0.008 | 15.530 | 0.006 | 14.651 | 0.012 | 9.235 | 8.469 | 7.590 | 0.427 |
| 4326.7984 | 4326.8003 | 4326.8023 | 16.248 | 0.010 | 15.513 | 0.009 | 14.661 | 0.018 | 9.187 | 8.452 | 7.600 | 0.480 |
| 4327.7919 | 4327.7939 | 4327.7958 | 16.355 | 0.016 | 15.626 | 0.019 | 14.738 | 0.019 | 9.294 | 8.565 | 7.677 | 0.672 |
| 4329.5381 | 4329.5401 | 4329.5421 | 16.267 | 0.007 | 15.536 | 0.005 | 14.779 | 0.004 | 9.207 | 8.476 | 7.719 | 1.012 |
| 4329.8096 | 4329.8116 | 4329.8135 | 16.272 | 0.007 | 15.546 | 0.007 | 14.776 | 0.009 | 9.212 | 8.486 | 7.716 | 1.065 |
| 4330.5588 | 4330.5607 | 4330.5627 | 16.260 | 0.007 | 15.533 | 0.005 | 14.666 | 0.007 | 9.199 | 8.472 | 7.605 | 1.210 |
| 4330.8044 | 4330.8063 | 4330.8083 | 16.329 | 0.006 | 15.647 | 0.006 | 14.795 | 0.005 | 9.268 | 8.586 | 7.734 | 1.258 |
| 4331.7645 | 4331.7665 | 4331.7685 | 16.317 | 0.006 | 15.607 | 0.006 | 14.707 | 0.010 | 9.256 | 8.546 | 7.646 | 1.444 |
| 4332.7703 | 4332.7722 | 4332.7741 | 16.368 | 0.022 | 15.615 | 0.018 | 14.691 | 0.013 | 9.307 | 8.554 | 7.630 | 1.639 |
| 4332.8366 | 4332.8386 | 4332.8405 | 16.315 | 0.050 | 15.536 | 0.037 | 14.657 | 0.023 | 9.254 | 8.475 | 7.596 | 1.652 |
| 4333.6676 | 4333.6696 | 4333.6715 | 16.355 | 0.005 | 15.633 | 0.004 | 14.733 | 0.019 | 9.293 | 8.571 | 7.671 | 1.812 |
| 4334.7252 | 4334.7272 | 4334.7291 | 16.399 | 0.007 | 15.660 | 0.006 | 14.766 | 0.030 | 9.336 | 8.597 | 7.703 | 2.016 |
| 4335.7036 | 4335.7056 | 4335.7075 | 16.377 | 0.007 | 15.606 | 0.006 | 14.682 | 0.006 | 9.313 | 8.542 | 7.618 | 2.203 |
| 4336.6460 | 4336.6480 | 4336.6499 | 16.360 | 0.008 | 15.562 | 0.011 | 14.638 | 0.016 | 9.295 | 8.497 | 7.573 | 2.383 |
| 4340.7052 | 4340.7072 | 4340.7092 | 16.470 | 0.050 | 15.780 | 0.049 | 14.930 | 0.014 | 9.400 | 8.710 | 7.860 | 3.146 |
| 4341.7061 | 4341.7081 | 4341.7101 | 16.440 | 0.012 | 15.663 | 0.009 | 14.700 | 0.014 | 9.368 | 8.591 | 7.628 | 3.332 |
| 4342.7469 | 4342.7489 | 4342.7508 | 16.605 | 0.014 | 15.805 | 0.012 | 14.899 | 0.013 | 9.531 | 8.731 | 7.825 | 3.523 |
| 4344.7438 | 4344.7459 | 4344.7478 | 16.439 | 0.012 | 15.724 | 0.006 | 14.863 | 0.008 | 9.361 | 8.646 | 7.785 | 3.885 |
| 4346.6986 | 4346.7006 | 4346.7025 | 16.425 | 0.017 | 15.685 | 0.027 | 14.766 | 0.031 | 9.343 | 8.603 | 7.684 | 4.234 |
| 4349.6979 | 4349.6999 | 4349.7018 | 16.549 | 0.008 | 15.805 | 0.005 | 14.877 | 0.008 | 9.459 | 8.715 | 7.787 | 4.757 |
| 4351.7302 | 4351.7322 | 4351.7341 | 16.474 | 0.007 | 15.768 | 0.005 | 14.800 | 0.023 | 9.379 | 8.673 | 7.704 | 5.102 |
| 4353.7300 | 4353.7320 | 4353.7339 | 16.547 | 0.008 | 15.814 | 0.009 | 14.878 | 0.005 | 9.445 | 8.712 | 7.776 | 5.433 |
| 4355.7561 | 4355.7581 | 4355.7601 | 16.620 | 0.038 | 15.866 | 0.043 | 14.986 | 0.042 | 9.512 | 8.758 | 7.878 | 5.760 |
| 4357.6853 | 4357.6873 | 4357.6893 | 16.554 | 0.008 | 15.826 | 0.005 | 14.972 | 0.005 | 9.439 | 8.711 | 7.857 | 6.062 |
| 4359.6567 | 4359.6587 | 4359.6607 | 16.563 | 0.008 | 15.840 | 0.008 | 14.905 | 0.009 | 9.440 | 8.717 | 7.782 | 6.362 |
| 4363.6516 | 4363.6536 | 4363.6555 | 16.622 | 0.008 | 15.912 | 0.010 | 15.008 | 0.007 | 9.483 | 8.773 | 7.869 | 6.941 |
| 4371.6303 | 4371.6323 | 4371.6343 | 16.725 | 0.009 | 16.051 | 0.006 | 15.077 | 0.009 | 9.551 | 8.877 | 7.903 | 7.969 |
| 4373.6526 | 4373.6546 | 4373.6565 | 16.735 | 0.007 | 15.997 | 0.005 | 15.066 | 0.006 | 9.551 | 8.813 | 7.882 | 8.202 |

### (b) 1208 Troilus

| JD (B) | JD (V) | JD (I) | B | $\sigma_B$ | V | $\sigma_V$ | I | $\sigma_I$ | $B_{red}$ | $V_{red}$ | $I_{red}$ | Phase |
|---|---|---|---|---|---|---|---|---|---|---|---|---|
| 4591.7622 | 4591.7642 | 4591.7663 | 16.681 | 0.013 | 15.989 | 0.007 | 15.265 | 0.006 | 9.793 | 9.101 | 8.377 | 0.625 |
| ... | 4592.7398 | 4592.7419 | ... | ... | 16.060 | 0.006 | 15.307 | 0.017 | ... | 9.173 | 8.420 | 0.427 |
| 4593.5733 | 4593.5753 | 4593.5774 | 16.728 | 0.007 | 15.984 | 0.005 | 15.255 | 0.005 | 9.841 | 9.097 | 8.368 | 0.264 |
| 4595.5365 | 4595.5386 | 4595.5406 | 16.597 | 0.107 | 15.928 | 0.005 | 15.170 | 0.005 | 9.712 | 9.043 | 8.285 | 0.199 |
| 4596.5263 | 4596.5283 | 4596.5304 | 16.712 | 0.006 | 15.994 | 0.005 | 15.251 | 0.005 | 9.827 | 9.109 | 8.366 | 0.387 |
| 4596.7865 | 4596.7885 | 4596.7906 | 16.772 | 0.015 | 16.073 | 0.005 | 15.201 | 0.006 | 9.887 | 9.188 | 8.316 | 0.439 |
| 4598.5141 | 4598.5161 | 4598.5181 | 16.679 | 0.015 | 15.974 | 0.008 | 15.208 | 0.013 | 9.795 | 9.090 | 8.324 | 0.791 |
| 4598.7825 | 4598.7845 | 4598.7866 | 16.670 | 0.009 | 15.977 | 0.006 | 15.207 | 0.006 | 9.786 | 9.093 | 8.323 | 0.846 |
| 4599.5063 | 4599.5083 | 4599.5103 | 16.797 | 0.009 | 16.048 | 0.009 | 15.292 | 0.013 | 9.913 | 9.164 | 8.408 | 0.995 |
| 4599.8075 | 4599.8095 | 4599.8116 | 16.779 | 0.012 | 16.012 | 0.010 | 15.240 | 0.014 | 9.895 | 9.128 | 8.356 | 1.057 |
| 4601.6080 | 4601.6100 | 4601.6121 | 16.750 | 0.006 | 16.101 | 0.005 | 15.326 | 0.006 | 9.865 | 9.216 | 8.441 | 1.426 |
| 4610.6473 | 4610.6493 | 4610.6514 | 16.812 | 0.008 | 16.123 | 0.008 | 15.379 | 0.010 | 9.918 | 9.229 | 8.485 | 3.247 |

### (c) 4348 Poulydamas

| JD (B) | JD (V) | JD (I) | B | $\sigma_B$ | V | $\sigma_V$ | I | $\sigma_I$ | $B_{red}$ | $V_{red}$ | $I_{red}$ | Phase |
|---|---|---|---|---|---|---|---|---|---|---|---|---|
| 4547.8036 | 4547.8056 | 4547.8077 | 18.228 | 0.028 | 17.540 | 0.017 | 16.725 | 0.018 | 10.961 | 10.273 | 9.458 | 6.821 |
| 4574.6735 | 4574.6755 | 4574.6776 | 17.866 | 0.016 | 17.179 | 0.012 | 16.430 | 0.011 | 10.699 | 10.012 | 9.263 | 2.438 |
| 4579.6968 | 4579.6988 | 4579.7009 | 17.599 | 0.012 | 16.954 | 0.012 | 16.234 | 0.009 | 10.441 | 9.796 | 9.076 | 1.493 |
| 4580.7074 | 4580.7094 | 4580.7115 | 17.599 | 0.042 | 17.017 | 0.046 | 16.219 | 0.037 | 10.442 | 9.860 | 9.062 | 1.300 |
| 4588.5336 | 4588.5356 | 4588.5377 | 17.517 | 0.011 | 16.961 | 0.009 | 16.167 | 0.010 | 10.366 | 9.810 | 9.016 | 0.207 |
| 4591.7355 | 4591.7375 | 4591.7396 | 17.679 | 0.009 | 16.987 | 0.007 | 16.218 | 0.008 | 10.528 | 9.836 | 9.067 | 0.824 |
| ... | 4592.7120 | 4592.7140 | ... | ... | 16.971 | 0.010 | 16.173 | 0.011 | ... | 9.819 | 9.021 | 1.012 |
| 4593.6761 | 4593.6781 | 4593.6802 | 17.929 | 0.018 | 17.176 | 0.011 | 16.387 | 0.013 | 10.777 | 10.024 | 9.235 | 1.197 |
| 4599.6463 | 4599.6483 | 4599.6504 | 18.011 | 0.012 | 17.260 | 0.010 | 16.482 | 0.011 | 10.853 | 10.102 | 9.324 | 2.330 |

### (d) 6998 Tithonus

| JD (B) | JD (V) | JD (I) | B | $\sigma_B$ | V | $\sigma_V$ | I | $\sigma_I$ | $B_{red}$ | $V_{red}$ | $I_{red}$ | Phase |
|---|---|---|---|---|---|---|---|---|---|---|---|---|
| 4567.7462 | 4567.7482 | 4567.7502 | 19.138 | 0.020 | 18.385 | 0.017 | 17.492 | 0.018 | 12.491 | 11.738 | 10.845 | 1.519 |
| 4569.5657 | 4569.5677 | 4569.5698 | 19.044 | 0.040 | 18.380 | 0.030 | 17.456 | 0.027 | 12.400 | 11.736 | 10.812 | 1.129 |
| 4569.7751 | 4569.7771 | 4569.7792 | 19.095 | 0.035 | 18.354 | 0.033 | 17.487 | 0.027 | 12.451 | 11.710 | 10.843 | 1.084 |
| 4574.5267 | 4574.5288 | 4574.5308 | 19.089 | 0.097 | 18.250 | 0.061 | 17.370 | 0.042 | 12.450 | 11.611 | 10.731 | 0.075 |
| 4574.8414 | 4574.8434 | 4574.8455 | 19.113 | 0.088 | 18.244 | 0.076 | 17.485 | 0.062 | 12.474 | 11.605 | 10.846 | 0.052 |
| 4579.5537 | ... | ... | 19.321 | 0.062 | ... | ... | ... | ... | 12.684 | ... | ... | 1.032 |
| 4579.8146 | 4579.8167 | 4579.5578 | 19.182 | 0.069 | 18.408 | 0.044 | 17.608 | 0.035 | 12.545 | 11.771 | 10.971 | 1.089 |
| 4580.6688 | 4580.6708 | 4580.6729 | 19.265 | 0.044 | 18.532 | 0.040 | 17.625 | 0.022 | 12.627 | 11.894 | 10.987 | 1.273 |
| 4588.6044 | 4588.6064 | 4588.6085 | 19.258 | 0.025 | 18.551 | 0.021 | 17.576 | 0.031 | 12.613 | 11.906 | 10.931 | 2.958 |
| 4591.5859 | 4591.5880 | 4591.5900 | 19.382 | 0.023 | 18.614 | 0.018 | 17.630 | 0.018 | 12.732 | 11.964 | 10.980 | 3.574 |
| 4596.5898 | 4596.5918 | 4596.5939 | 19.404 | 0.022 | 18.578 | 0.017 | 17.635 | 0.018 | 12.743 | 11.917 | 10.974 | 4.576 |

**(e) 8317 Eurysaces**

| JD (B) | JD (V) | JD (I) | B | $\sigma_B$ | V | $\sigma_V$ | I | $\sigma_I$ | $B_{red}$ | $V_{red}$ | $I_{red}$ | Phase |
|--------|--------|--------|---|------------|---|------------|---|------------|-----------|-----------|-----------|-------|
| 3899.8754 | ... | ... | 19.542 | 0.141 | ... | ... | ... | ... | 12.783 | ... | ... | 10.377 |
| 3899.8782 | 3899.8809 | 3899.8823 | 19.349 | 0.113 | 18.595 | 0.119 | 17.777 | 0.110 | 12.590 | 11.836 | 11.018 | 10.377 |
| 3900.8715 | ... | ... | 19.521 | 0.063 | ... | ... | ... | ... | 12.768 | ... | ... | 10.287 |
| 3900.8742 | 3900.8770 | 3900.8784 | 19.527 | 0.061 | 18.733 | 0.062 | 17.768 | 0.048 | 12.774 | 11.980 | 11.015 | 10.287 |
| 3901.8312 | ... | ... | 19.162 | 0.102 | ... | ... | ... | ... | 12.416 | ... | ... | 10.198 |
| 3901.8339 | 3901.8367 | 3901.8381 | 19.206 | 0.110 | 18.480 | 0.104 | 17.973 | 0.147 | 12.460 | 11.734 | 11.227 | 10.198 |
| 3903.8343 | ... | ... | 19.367 | 0.061 | ... | ... | ... | ... | 12.634 | ... | ... | 10.003 |
| 3903.8370 | 3903.8398 | 3903.8412 | 19.468 | 0.067 | 18.572 | 0.076 | 17.711 | 0.075 | 12.735 | 11.839 | 10.978 | 10.003 |
| 3905.7776 | ... | ... | 19.437 | 0.036 | ... | ... | ... | ... | 12.716 | ... | ... | 9.801 |
| 3905.7803 | 3905.7831 | 3905.7845 | 19.413 | 0.032 | 18.623 | 0.044 | 17.896 | 0.054 | 12.692 | 11.902 | 11.175 | 9.801 |
| 3907.7848 | ... | ... | 19.391 | 0.025 | ... | ... | ... | ... | 12.683 | ... | ... | 9.580 |
| 3907.7875 | 3907.7903 | 3907.7917 | 19.388 | 0.024 | 18.698 | 0.037 | 17.785 | 0.045 | 12.680 | 11.990 | 11.077 | 9.579 |
| 3909.8098 | ... | ... | 19.457 | 0.024 | ... | ... | ... | ... | 12.762 | ... | ... | 9.343 |
| 3909.8125 | 3909.8152 | 3909.8166 | 19.473 | 0.024 | 18.817 | 0.039 | 17.902 | 0.038 | 12.778 | 12.122 | 11.207 | 9.343 |
| 3912.8506 | ... | ... | 19.293 | 0.023 | ... | ... | ... | ... | 12.616 | ... | ... | 8.963 |
| 3912.8533 | 3912.8560 | 3912.8574 | 19.254 | 0.021 | 18.538 | 0.032 | 17.662 | 0.035 | 12.577 | 11.861 | 10.985 | 8.963 |
| 3916.8480 | ... | ... | 19.172 | 0.031 | ... | ... | ... | ... | 12.519 | ... | ... | 8.420 |
| 3916.8507 | 3916.8535 | 3916.8549 | 19.144 | 0.032 | 18.529 | 0.064 | 17.678 | 0.055 | 12.491 | 11.876 | 11.025 | 8.420 |
| 3935.7560 | ... | ... | 19.120 | 0.025 | ... | ... | ... | ... | 12.556 | ... | ... | 5.238 |
| 3935.7587 | 3935.7615 | 3935.7629 | 19.153 | 0.022 | 18.416 | 0.038 | 17.554 | 0.034 | 12.589 | 11.852 | 10.990 | 5.237 |
| 3937.7593 | ... | ... | 19.100 | 0.018 | ... | ... | ... | ... | 12.543 | ... | ... | 4.849 |
| 3937.7620 | 3937.7648 | 3937.7662 | 19.111 | 0.020 | 18.280 | 0.027 | 17.377 | 0.028 | 12.554 | 11.723 | 10.820 | 4.849 |
| 3945.7493 | ... | ... | 18.993 | 0.024 | ... | ... | ... | ... | 12.459 | ... | ... | 3.226 |
| 3945.7520 | 3945.7548 | 3945.7562 | 19.023 | 0.023 | 18.206 | 0.029 | 17.340 | 0.029 | 12.489 | 11.672 | 10.806 | 3.226 |
| 3951.8935 | ... | ... | 18.525 | 0.154 | ... | ... | ... | ... | 12.002 | ... | ... | 1.920 |
| 3951.8935 | 3951.8935 | 3951.8935 | 18.674 | 0.126 | 18.053 | 0.148 | 17.088 | 0.084 | 12.151 | 11.530 | 10.565 | 1.920 |
| 3958.7068 | ... | ... | 18.691 | 0.041 | ... | ... | ... | ... | 12.173 | ... | ... | 0.474 |
| 3958.7095 | 3958.7122 | 3958.7136 | 18.601 | 0.043 | 17.986 | 0.049 | 17.066 | 0.042 | 12.083 | 11.468 | 10.548 | 0.474 |
| 3960.7681 | ... | ... | 18.609 | 0.023 | ... | ... | ... | ... | 12.091 | ... | ... | 0.225 |
| 3960.7708 | 3960.7736 | 3960.7750 | 18.698 | 0.024 | 17.980 | 0.030 | 17.116 | 0.027 | 12.180 | 11.462 | 10.598 | 0.225 |
| 3964.7720 | ... | ... | 18.623 | 0.030 | ... | ... | ... | ... | 12.103 | ... | ... | 0.940 |
| 3964.7747 | 3964.7775 | 3964.7789 | 18.636 | 0.028 | 17.874 | 0.023 | 16.989 | 0.026 | 12.116 | 11.354 | 10.469 | 0.941 |
| 3966.7712 | ... | ... | 18.640 | 0.017 | ... | ... | ... | ... | 12.119 | ... | ... | 1.369 |
| 3966.7740 | 3966.7767 | 3966.7781 | 18.632 | 0.017 | 17.853 | 0.024 | 17.100 | 0.025 | 12.111 | 11.332 | 10.579 | 1.370 |
| 3968.7893 | ... | ... | 18.764 | 0.018 | ... | ... | ... | ... | 12.240 | ... | ... | 1.804 |
| 3968.7921 | 3968.7948 | 3968.7962 | 18.824 | 0.018 | 17.990 | 0.026 | 17.059 | 0.028 | 12.300 | 11.466 | 10.535 | 1.804 |
| 3975.7384 | ... | ... | 18.892 | 0.019 | ... | ... | ... | ... | 12.355 | ... | ... | 3.279 |
| 3975.7411 | 3975.7439 | 3975.7453 | 18.840 | 0.020 | 18.071 | 0.034 | 17.203 | 0.031 | 12.302 | 11.533 | 10.665 | 3.280 |
| 3998.6164 | ... | ... | 19.275 | 0.025 | ... | ... | ... | ... | 12.646 | ... | ... | 7.528 |
| 3998.6191 | 3998.6219 | 3998.6233 | 19.232 | 0.025 | 18.446 | 0.034 | 17.597 | 0.038 | 12.603 | 11.817 | 10.968 | 7.528 |
| 4292.8440 | ... | ... | 19.463 | 0.043 | ... | ... | ... | ... | 12.613 | ... | ... | 11.067 |
| 4292.8467 | ... | 4292.8507 | 19.518 | 0.044 | ... | ... | 17.887 | 0.079 | 12.668 | ... | 11.037 | 11.066 |
| 4310.8524 | 4310.8564 | ... | 19.262 | 0.061 | 18.595 | 0.044 | ... | ... | 12.528 | 11.861 | ... | 9.499 |
| 4316.8476 | ... | ... | 19.250 | 0.039 | ... | ... | ... | ... | 12.551 | ... | ... | 8.748 |

**(e)(cont.) 8317 Eurysaces**

| JD (B) | JD (V) | JD (I) | B | $\sigma_B$ | V | $\sigma_V$ | I | $\sigma_I$ | $B_{red}$ | $V_{red}$ | $I_{red}$ | Phase |
|---|---|---|---|---|---|---|---|---|---|---|---|---|
| 4316.8502 | 4316.8529 | 4316.8543 | 19.236 | 0.044 | 18.464 | 0.049 | 17.627 | 0.044 | 12.537 | 11.765 | 10.928 | 8.748 |
| 4331.8552 | ... | ... | 19.088 | 0.022 | ... | ... | ... | ... | 12.463 | ... | ... | 6.394 |
| 4331.8579 | 4331.8606 | 4331.8619 | 19.146 | 0.022 | 18.368 | 0.038 | 17.432 | 0.041 | 12.522 | 11.744 | 10.808 | 6.393 |
| 4331.8661 | ... | ... | 19.096 | 0.024 | ... | ... | ... | ... | 12.472 | ... | ... | 6.392 |
| 4331.8688 | 4331.8715 | 4331.8728 | 19.119 | 0.025 | 18.367 | 0.037 | 17.440 | 0.038 | 12.495 | 11.743 | 10.816 | 6.391 |
| 4333.7448 | ... | ... | 18.986 | 0.019 | ... | ... | ... | ... | 12.369 | ... | ... | 6.054 |
| 4333.7475 | 4333.7502 | 4333.7515 | 19.029 | 0.026 | 18.270 | 0.028 | 17.382 | 0.032 | 12.412 | 11.653 | 10.765 | 6.053 |
| 4335.8654 | ... | ... | 19.102 | 0.028 | ... | ... | ... | ... | 12.493 | ... | ... | 5.662 |
| 4335.8681 | 4335.8708 | 4335.8721 | 19.089 | 0.030 | 18.262 | 0.045 | 17.449 | 0.041 | 12.480 | 11.653 | 10.840 | 5.662 |
| 4338.7836 | 4338.7876 | 4338.7856 | 19.084 | 0.055 | 18.268 | 0.035 | 17.419 | 0.027 | 12.486 | 11.670 | 10.821 | 5.106 |
| 4339.8521 | 4339.8562 | 4339.8542 | 18.827 | 0.067 | 18.317 | 0.045 | 17.278 | 0.236 | 12.232 | 11.722 | 10.683 | 4.898 |
| 4340.7436 | 4340.7476 | 4340.7456 | 18.860 | 0.095 | 18.140 | 0.046 | 17.349 | 0.052 | 12.268 | 11.548 | 10.757 | 4.723 |
| 4345.8049 | ... | ... | 18.920 | 0.031 | ... | ... | ... | ... | 12.342 | ... | ... | 3.701 |
| 4345.8075 | 4345.8102 | 4345.8116 | 18.885 | 0.030 | 18.187 | 0.038 | 17.378 | 0.034 | 12.307 | 11.609 | 10.800 | 3.700 |
| 4348.7449 | ... | ... | 18.828 | 0.020 | ... | ... | ... | ... | 12.256 | ... | ... | 3.088 |
| 4348.7476 | 4348.7503 | 4348.7516 | 18.822 | 0.016 | 18.075 | 0.025 | 17.252 | 0.032 | 12.250 | 11.503 | 10.680 | 3.087 |
| 4350.8353 | 4350.8394 | 4350.8374 | 18.940 | 0.024 | 18.147 | 0.020 | 17.312 | 0.019 | 12.372 | 11.579 | 10.744 | 2.645 |
| 4351.7786 | 4351.7826 | 4351.7806 | 18.755 | 0.023 | 18.139 | 0.024 | 17.308 | 0.018 | 12.188 | 11.572 | 10.741 | 2.443 |
| 4352.8022 | 4352.8063 | 4352.8043 | 18.964 | 0.038 | 18.159 | 0.021 | 17.297 | 0.024 | 12.398 | 11.593 | 10.731 | 2.223 |
| 4353.7999 | 4353.8040 | 4353.8020 | 18.719 | 0.038 | 18.034 | 0.034 | 17.142 | 0.032 | 12.155 | 11.470 | 10.578 | 2.008 |
| 4354.8471 | 4354.8511 | 4354.8491 | 18.813 | 0.025 | 18.108 | 0.021 | 17.253 | 0.045 | 12.250 | 11.545 | 10.690 | 1.782 |
| 4355.8013 | 4355.8054 | 4355.8034 | 18.659 | 0.019 | 17.927 | 0.015 | 17.016 | 0.019 | 12.097 | 11.365 | 10.454 | 1.575 |
| 4356.7567 | ... | ... | 18.755 | 0.022 | ... | ... | ... | ... | 12.193 | ... | ... | 1.367 |
| 4356.7593 | 4356.7620 | 4356.7634 | 18.730 | 0.020 | 18.047 | 0.033 | 17.256 | 0.025 | 12.168 | 11.485 | 10.694 | 1.367 |
| 4357.7522 | ... | ... | 18.687 | 0.016 | ... | ... | ... | ... | 12.126 | ... | ... | 1.151 |
| 4357.7548 | 4357.7575 | 4357.7588 | 18.702 | 0.015 | 17.935 | 0.024 | 17.172 | 0.024 | 12.141 | 11.374 | 10.611 | 1.150 |
| 4358.5742 | ... | ... | 18.616 | 0.024 | ... | ... | ... | ... | 12.055 | ... | ... | 0.972 |
| 4358.5769 | 4358.5796 | 4358.5810 | 18.627 | 0.021 | 18.027 | 0.035 | 17.257 | 0.033 | 12.066 | 11.466 | 10.696 | 0.971 |
| 4358.8000 | ... | ... | 18.641 | 0.022 | ... | ... | ... | ... | 12.080 | ... | ... | 0.923 |
| 4358.8027 | 4358.8054 | 4358.8067 | 18.592 | 0.022 | 17.881 | 0.029 | 17.175 | 0.030 | 12.031 | 11.320 | 10.614 | 0.922 |
| 4359.5559 | ... | ... | 18.473 | 0.031 | ... | ... | ... | ... | 11.912 | ... | ... | 0.759 |
| 4359.5586 | 4359.5613 | 4359.5626 | 18.495 | 0.036 | 17.940 | 0.032 | 17.000 | 0.037 | 11.934 | 11.379 | 10.439 | 0.758 |
| 4359.7985 | ... | ... | 18.513 | 0.025 | ... | ... | ... | ... | 11.952 | ... | ... | 0.706 |
| 4359.8011 | 4359.8038 | 4359.8052 | 18.658 | 0.024 | 17.949 | 0.035 | 17.134 | 0.045 | 12.097 | 11.388 | 10.573 | 0.705 |
| 4360.6728 | ... | ... | 18.649 | 0.037 | ... | ... | ... | ... | 12.089 | ... | ... | 0.518 |
| 4360.8412 | ... | ... | 18.704 | 0.028 | ... | ... | ... | ... | 12.144 | ... | ... | 0.482 |
| 4360.8439 | 4360.8466 | 4360.8480 | 18.678 | 0.026 | 17.989 | 0.036 | 17.223 | 0.040 | 12.118 | 11.429 | 10.663 | 0.482 |
| 4361.6509 | 4361.6550 | 4361.6530 | 18.701 | 0.027 | 17.974 | 0.025 | 17.076 | 0.024 | 12.141 | 11.414 | 10.516 | 0.314 |
| 4361.8054 | 4361.8095 | 4361.8075 | 18.756 | 0.029 | 18.015 | 0.023 | 17.175 | 0.023 | 12.195 | 11.454 | 10.614 | 0.283 |
| 4362.5619 | 4362.5660 | 4362.5640 | 18.565 | 0.028 | 17.957 | 0.025 | 17.024 | 0.026 | 12.004 | 11.396 | 10.463 | 0.153 |
| 4362.7682 | 4362.7723 | 4362.7702 | 18.547 | 0.027 | 18.014 | 0.031 | 17.106 | 0.026 | 11.986 | 11.453 | 10.545 | 0.132 |
| 4363.5534 | 4363.5574 | 4363.5554 | 18.709 | 0.177 | 18.018 | 0.041 | 16.975 | 0.106 | 12.148 | 11.457 | 10.414 | 0.181 |
| 4363.7295 | 4363.7336 | 4363.7316 | 18.698 | 0.030 | 18.107 | 0.022 | 17.173 | 0.023 | 12.137 | 11.546 | 10.612 | 0.211 |
| 4364.5818 | 4364.5859 | 4364.5838 | 18.675 | 0.044 | 18.077 | 0.030 | 17.206 | 0.025 | 12.113 | 11.515 | 10.644 | 0.378 |

**(e)(cont.)  8317 Eurysaces**

| JD (B) | JD (V) | JD (I) | B | $\sigma_B$ | V | $\sigma_V$ | I | $\sigma_I$ | $B_{red}$ | $V_{red}$ | $I_{red}$ | Phase |
|--------|--------|--------|---|-----------|---|-----------|---|-----------|-----------|-----------|-----------|-------|
| 4365.5468 | 4365.5509 | 4365.5488 | 18.713 | 0.100 | 18.029 | 0.059 | 17.105 | 0.085 | 12.151 | 11.467 | 10.543 | 0.583 |
| 4366.8067 | ... | 4366.8087 | 18.601 | 0.088 | ... | ... | 17.155 | 0.040 | 12.038 | ... | 10.592 | 0.855 |
| 4367.5424 | 4367.5465 | 4367.5444 | 18.604 | 0.060 | 18.058 | 0.039 | 17.197 | 0.033 | 12.040 | 11.494 | 10.633 | 1.014 |
| 4367.7876 | 4367.7916 | 4367.7896 | 19.071 | 0.184 | 18.121 | 0.084 | 17.258 | 0.095 | 12.507 | 11.557 | 10.694 | 1.067 |
| 4368.8051 | 4368.8092 | 4368.8071 | 18.646 | 0.137 | 18.089 | 0.104 | 17.102 | 0.064 | 12.080 | 11.523 | 10.536 | 1.288 |
| 4371.7150 | ... | ... | 18.641 | 0.048 | ... | ... | ... | ... | 12.071 | ... | ... | 1.918 |
| 4371.7177 | 4371.7205 | 4371.7218 | 18.216 | 0.236 | 18.016 | 0.061 | 17.118 | 0.041 | 11.646 | 11.446 | 10.548 | 1.918 |
| 4372.7473 | ... | ... | 18.594 | 0.030 | ... | ... | ... | ... | 12.022 | ... | ... | 2.139 |
| 4372.7499 | 4372.7527 | 4372.7540 | 18.704 | 0.049 | 17.985 | 0.047 | 17.149 | 0.036 | 12.132 | 11.413 | 10.577 | 2.140 |
| 4372.7572 | ... | ... | 18.795 | 0.034 | ... | ... | ... | ... | 12.223 | ... | ... | 2.142 |
| 4372.7598 | 4372.7626 | 4372.7640 | 18.786 | 0.035 | 18.097 | 0.039 | 17.227 | 0.031 | 12.214 | 11.525 | 10.655 | 2.142 |
| 4373.7274 | ... | ... | 18.842 | 0.031 | ... | ... | ... | ... | 12.268 | ... | ... | 2.349 |
| 4373.7301 | 4373.7328 | 4373.7342 | 18.824 | 0.032 | 18.081 | 0.032 | 17.246 | 0.030 | 12.250 | 11.507 | 10.672 | 2.350 |
| 4374.7895 | ... | ... | 18.772 | 0.067 | ... | ... | ... | ... | 12.196 | ... | ... | 2.575 |
| 4374.7922 | 4374.7949 | 4374.7963 | 18.821 | 0.094 | 18.108 | 0.075 | 17.234 | 0.057 | 12.245 | 11.532 | 10.658 | 2.576 |
| 4375.6841 | ... | ... | 18.918 | 0.028 | ... | ... | ... | ... | 12.340 | ... | ... | 2.765 |
| 4375.6867 | 4375.6895 | 4375.6908 | 18.985 | 0.034 | 18.063 | 0.044 | 17.308 | 0.045 | 12.407 | 11.485 | 10.730 | 2.765 |

**(f) 12126 (1999 RM11)**

| JD (B) | JD (V) | JD (I) | B | $\sigma_B$ | V | $\sigma_V$ | I | $\sigma_I$ | $B_{red}$ | $V_{red}$ | $I_{red}$ | Phase |
|--------|--------|--------|---|-----------|---|-----------|---|-----------|-----------|-----------|-----------|-------|
| 4551.7874 | 4551.7895 | 4551.7915 | 17.955 | 0.022 | 17.272 | 0.030 | 16.390 | 0.019 | 11.173 | 10.490 | 9.608 | 0.956 |
| 4554.6254 | 4554.6274 | 4554.6295 | 17.977 | 0.014 | 17.289 | 0.010 | 16.332 | 0.011 | 11.201 | 10.513 | 9.556 | 0.379 |
| 4554.8191 | 4554.8211 | 4554.8232 | 17.999 | 0.014 | 17.320 | 0.013 | 16.427 | 0.011 | 11.223 | 10.544 | 9.651 | 0.342 |
| 4555.5536 | 4555.5556 | 4555.5576 | 17.847 | 0.012 | 17.153 | 0.011 | 16.236 | 0.010 | 11.072 | 10.378 | 9.461 | 0.216 |
| 4555.8399 | 4555.8420 | 4555.8440 | 17.911 | 0.017 | 17.238 | 0.011 | 16.339 | 0.012 | 11.137 | 10.464 | 9.565 | 0.180 |
| 4556.5877 | 4556.5897 | 4556.5918 | 17.845 | 0.011 | 17.231 | 0.009 | 16.331 | 0.011 | 11.072 | 10.458 | 9.558 | 0.173 |
| 4556.7671 | 4556.7691 | 4556.7711 | 18.020 | 0.011 | 17.334 | 0.009 | 16.415 | 0.009 | 11.247 | 10.561 | 9.642 | 0.191 |
| 4557.5428 | 4557.5448 | 4557.5468 | 17.770 | 0.013 | 17.101 | 0.017 | 16.197 | 0.012 | 10.998 | 10.329 | 9.425 | 0.315 |
| 4557.8198 | 4557.8218 | 4557.8239 | 17.956 | 0.013 | 17.249 | 0.017 | 16.318 | 0.010 | 11.185 | 10.478 | 9.547 | 0.367 |
| 4559.5405 | 4559.5425 | 4559.5446 | 17.790 | 0.013 | 17.268 | 0.014 | 16.180 | 0.027 | 11.021 | 10.499 | 9.411 | 0.714 |
| 4559.8295 | 4559.8315 | 4559.8335 | 17.961 | 0.013 | 17.357 | 0.011 | 16.357 | 0.020 | 11.192 | 10.588 | 9.588 | 0.774 |
| 4569.7281 | 4569.7302 | 4569.7322 | 17.943 | 0.012 | 17.274 | 0.010 | 16.358 | 0.012 | 11.177 | 10.508 | 9.592 | 2.835 |
| 4580.6337 | 4580.6357 | 4580.6378 | 18.122 | 0.018 | 17.394 | 0.019 | 16.487 | 0.017 | 11.342 | 10.614 | 9.707 | 4.989 |
| ... | ... | 4592.6605 | ... | ... | ... | ... | 16.787 | 0.012 | ... | ... | 9.974 | 7.096 |

## (g) 13323 (1998 SQ)

| JD (B) | JD (V) | JD (I) | B | $\sigma_B$ | V | $\sigma_V$ | I | $\sigma_I$ | $B_{red}$ | $V_{red}$ | $I_{red}$ | Phase |
|---|---|---|---|---|---|---|---|---|---|---|---|---|
| 4180.8705 | ... | ... | 20.089 | 0.081 | ... | ... | ... | ... | 12.496 | ... | ... | 9.362 |
| 4180.8725 | 4180.8744 | 4180.8764 | 20.073 | 0.078 | 19.213 | 0.058 | 18.431 | 0.063 | 12.480 | 11.620 | 10.838 | 9.362 |
| 4191.8325 | 4191.8344 | 4191.8364 | 19.840 | 0.262 | 19.085 | 0.119 | 18.436 | 0.110 | 12.311 | 11.556 | 10.907 | 9.975 |
| 4192.8612 | ... | ... | 19.776 | 0.161 | ... | ... | 18.411 | 0.066 | 12.254 | ... | ... | 10.019 |
| 4192.8631 | 4192.8651 | 4192.8671 | 19.923 | 0.167 | 19.318 | 0.110 | 18.411 | 0.066 | 12.401 | 11.796 | 10.889 | 10.020 |
| 4193.3647 | 4193.8667 | 4193.8687 | 19.917 | 0.161 | 19.277 | 0.095 | 18.383 | 0.062 | 12.401 | 11.761 | 10.867 | 10.061 |
| ... | 4194.8705 | ... | ... | ... | 19.189 | 0.078 | ... | ... | ... | 11.679 | ... | 10.099 |
| ... | 4194.8724 | 4194.8744 | ... | ... | 19.165 | 0.078 | 18.293 | 0.055 | ... | 11.655 | 10.783 | 10.100 |
| ... | 4196.8782 | ... | ... | ... | 19.431 | 0.080 | ... | ... | ... | 11.934 | ... | 10.170 |
| ... | 4196.8802 | 4196.8822 | ... | ... | 19.257 | 0.067 | 18.601 | 0.069 | ... | 11.760 | 11.104 | 10.170 |
| ... | 4197.8703 | ... | ... | ... | 19.294 | 0.064 | ... | ... | ... | 11.803 | ... | 10.202 |
| ... | 4197.8723 | 4197.8743 | ... | ... | 19.345 | 0.061 | 18.332 | 0.051 | ... | 11.854 | 10.841 | 10.202 |
| ... | 4198.8693 | ... | ... | ... | 19.247 | 0.058 | ... | ... | ... | 11.762 | ... | 10.231 |
| ... | 4198.8712 | 4198.8732 | ... | ... | 19.188 | 0.054 | 18.475 | 0.057 | ... | 11.703 | 10.990 | 10.231 |
| ... | 4216.7947 | ... | ... | ... | 19.097 | 0.033 | ... | ... | ... | 11.730 | ... | 10.330 |
| ... | 4216.7967 | 4216.7987 | ... | ... | 19.082 | 0.036 | 18.253 | 0.036 | ... | 11.715 | 10.886 | 10.330 |
| ... | 4219.9127 | ... | ... | ... | 19.051 | 0.032 | ... | ... | ... | 11.705 | ... | 10.260 |
| ... | 4219.9147 | 4219.9167 | ... | ... | 19.054 | 0.031 | 18.243 | 0.034 | ... | 11.708 | 10.897 | 10.260 |
| ... | 4257.8482 | ... | ... | ... | 18.909 | 0.061 | ... | ... | ... | 11.800 | ... | 7.140 |
| ... | 4257.8502 | 4257.8522 | ... | ... | 18.860 | 0.065 | 17.999 | 0.063 | ... | 11.751 | 10.890 | 7.140 |
| ... | 4277.8159 | ... | ... | ... | 18.460 | 0.022 | ... | ... | ... | 11.436 | ... | 3.942 |
| ... | 4277.8179 | 4277.8199 | ... | ... | 18.488 | 0.022 | 17.623 | 0.023 | ... | 11.464 | 10.599 | 3.942 |
| 4293.8161 | ... | ... | 19.054 | 0.024 | ... | ... | ... | ... | 12.064 | ... | ... | 0.909 |
| 4293.8181 | 4293.8201 | 4293.8221 | 19.064 | 0.023 | 18.329 | 0.020 | 17.551 | 0.021 | 12.074 | 11.339 | 10.561 | 0.908 |
| ... | 4295.8227 | ... | ... | ... | 18.344 | 0.018 | ... | ... | ... | 11.356 | ... | 0.528 |
| ... | 4295.8246 | 4295.8267 | ... | ... | 18.312 | 0.018 | 17.518 | 0.021 | ... | 11.324 | 10.530 | 0.527 |
| ... | 4295.8455 | ... | ... | ... | 18.278 | 0.019 | ... | ... | ... | 11.290 | ... | 0.523 |
| ... | 4295.8474 | 4295.8494 | ... | ... | 18.334 | 0.020 | 17.533 | 0.022 | ... | 11.346 | 10.545 | 0.523 |
| ... | 4296.8077 | ... | ... | ... | 18.281 | 0.018 | ... | ... | ... | 11.293 | ... | 0.353 |
| ... | 4296.8097 | 4296.8117 | ... | ... | 18.266 | 0.018 | 17.500 | 0.019 | ... | 11.278 | 10.512 | 0.352 |
| ... | 4296.8255 | ... | ... | ... | 18.256 | 0.018 | ... | ... | ... | 11.268 | ... | 0.350 |
| ... | 4296.8274 | 4296.8295 | ... | ... | 18.262 | 0.020 | 17.439 | 0.017 | ... | 11.274 | 10.451 | 0.349 |
| ... | 4300.7132 | ... | ... | ... | 18.380 | 0.021 | ... | ... | ... | 11.394 | ... | 0.514 |
| ... | 4300.7152 | 4300.7172 | ... | ... | 18.390 | 0.021 | 17.533 | 0.023 | ... | 11.404 | 10.547 | 0.514 |
| ... | 4300.7408 | ... | ... | ... | 18.261 | 0.019 | ... | ... | ... | 11.275 | ... | 0.519 |
| ... | 4300.7427 | 4300.7448 | ... | ... | 18.311 | 0.021 | 17.443 | 0.024 | ... | 11.325 | 10.457 | 0.519 |
| 4301.7610 | ... | ... | 19.077 | 0.023 | ... | ... | ... | ... | 12.091 | ... | ... | 0.711 |
| 4301.7630 | 4301.7650 | 4301.7670 | 19.055 | 0.023 | 18.308 | 0.018 | 17.526 | 0.021 | 12.069 | 11.322 | 10.540 | 0.712 |
| 4301.7849 | ... | ... | 19.064 | 0.024 | ... | ... | ... | ... | 12.078 | ... | ... | 0.715 |
| 4301.7869 | 4301.7889 | 4301.7909 | 19.021 | 0.022 | 18.314 | 0.020 | 17.495 | 0.022 | 12.035 | 11.328 | 10.509 | 0.716 |
| 4304.7512 | ... | ... | 18.998 | 0.035 | ... | ... | ... | ... | 12.010 | ... | ... | 1.289 |
| 4304.7531 | 4304.7551 | 4304.7571 | 19.014 | 0.034 | 18.265 | 0.027 | 17.503 | 0.028 | 12.026 | 11.277 | 10.515 | 1.290 |
| 4305.7426 | ... | ... | 18.859 | 0.053 | ... | ... | ... | ... | 11.871 | ... | ... | 1.482 |
| 4305.7445 | 4305.7465 | ... | 18.824 | 0.054 | 18.040 | 0.041 | ... | ... | 11.836 | 11.052 | ... | 1.483 |

**(g)(cont.)  13323 (1998 SQ)**

| JD (B) | JD (V) | JD (I) | B | $\sigma_B$ | V | $\sigma_V$ | I | $\sigma_I$ | $B_{red}$ | $V_{red}$ | $I_{red}$ | Phase |
|---|---|---|---|---|---|---|---|---|---|---|---|---|
| 4307.7410 | ... | ... | 19.120 | 0.041 | ... | ... | ... | ... | 12.130 | ... | ... | 1.870 |
| 4307.7430 | 4307.7450 | 4307.7471 | 19.107 | 0.043 | 18.430 | 0.030 | 17.541 | 0.027 | 12.117 | 11.440 | 10.551 | 1.871 |
| ... | 4308.6800 | ... | ... | ... | 18.389 | 0.029 | ... | ... | ... | 11.398 | ... | 2.052 |
| ... | 4308.6819 | 4308.6839 | ... | ... | 18.414 | 0.033 | 17.610 | 0.032 | ... | 11.423 | 10.619 | 2.052 |
| ... | 4339.7266 | ... | ... | ... | 18.785 | 0.088 | ... | ... | ... | 11.700 | ... | 7.314 |
| ... | 4339.7286 | 4339.7307 | ... | ... | 18.616 | 0.077 | 17.851 | 0.064 | ... | 11.531 | 10.766 | 7.314 |
| ... | 4372.5875 | ... | ... | ... | 19.137 | 0.046 | ... | ... | ... | 11.868 | ... | 10.207 |
| ... | 4372.5895 | 4372.5915 | ... | ... | 19.067 | 0.044 | 18.226 | 0.040 | ... | 11.798 | 10.957 | 10.207 |
| ... | 4375.5717 | ... | ... | ... | 18.992 | 0.056 | ... | ... | ... | 11.705 | ... | 10.311 |
| ... | 4375.5737 | 4375.5758 | ... | ... | 19.024 | 0.066 | 18.139 | 0.083 | ... | 11.737 | 10.852 | 10.311 |

## (h) 24506 (2001 BS15)

| JD (B) | JD (V) | JD (I) | B | $\sigma_B$ | V | $\sigma_V$ | I | $\sigma_I$ | $B_{red}$ | $V_{red}$ | $I_{red}$ | Phase |
|---|---|---|---|---|---|---|---|---|---|---|---|---|
| 4293.8571 | ... | 4293.8611 | 19.376 | 0.030 | ... | | 17.671 | 0.022 | 12.339 | ... | 10.634 | 8.651 |
| 4315.7736 | 4315.7755 | 4315.7776 | 18.866 | 0.078 | 18.221 | 0.053 | 17.356 | 0.041 | 11.938 | 11.293 | 10.428 | 5.499 |
| 4319.7588 | 4319.7607 | 4319.7627 | 19.077 | 0.022 | 18.286 | 0.018 | 17.334 | 0.017 | 12.164 | 11.373 | 10.421 | 4.799 |
| ... | 4325.7825 | 4325.7847 | ... | | 18.008 | 0.069 | 17.079 | 0.081 | ... | 11.113 | 10.184 | 3.683 |
| 4329.7708 | 4329.7728 | 4329.7748 | 18.814 | 0.130 | 18.073 | 0.125 | 17.238 | 0.144 | 11.929 | 11.188 | 10.353 | 2.912 |
| 4331.8434 | 4331.8454 | 4331.8474 | 18.857 | 0.026 | 18.101 | 0.021 | 17.135 | 0.023 | 11.976 | 11.220 | 10.254 | 2.504 |
| ... | 4334.7534 | 4334.7554 | | | 18.058 | 0.017 | 17.119 | 0.019 | ... | 11.182 | 10.243 | 1.924 |
| 4336.7192 | 4336.7212 | 4336.7232 | 18.743 | 0.052 | 18.142 | 0.034 | 17.002 | 0.024 | 11.869 | 11.268 | 10.128 | 1.527 |
| 4338.7740 | 4338.7760 | 4338.7781 | 18.670 | 0.041 | 17.963 | 0.026 | 17.039 | 0.081 | 11.798 | 11.091 | 10.167 | 1.111 |
| 4339.8254 | 4339.8274 | 4339.8295 | 18.740 | 0.061 | 18.003 | 0.039 | 17.154 | 0.033 | 11.869 | 11.132 | 10.283 | 0.899 |
| 4342.5556 | ... | 4342.5597 | 19.245 | 0.097 | ... | | 17.583 | 0.024 | 12.375 | ... | 10.713 | 0.352 |
| ... | ... | 4342.8460 | ... | | ... | | 17.580 | 0.030 | ... | ... | 10.710 | 0.297 |
| 4343.5487 | 4343.5507 | 4343.5527 | 18.704 | 0.027 | 18.041 | 0.021 | 17.033 | 0.020 | 11.834 | 11.171 | 10.163 | 0.173 |
| 4343.8258 | 4343.8278 | 4343.8298 | 18.661 | 0.050 | 17.978 | 0.042 | 17.035 | 0.027 | 11.791 | 11.108 | 10.165 | 0.136 |
| 4344.7311 | 4344.7331 | 4344.7351 | 18.763 | 0.048 | 17.967 | 0.019 | 17.046 | 0.021 | 11.893 | 11.097 | 10.176 | 0.162 |
| 4344.8010 | 4344.8030 | ... | 18.705 | 0.035 | 18.027 | 0.025 | ... | | 11.835 | 11.157 | ... | 0.172 |
| 4345.6842 | 4345.6862 | 4345.6882 | 18.648 | 0.025 | 17.905 | 0.019 | 17.022 | 0.019 | 11.778 | 11.035 | 10.152 | 0.329 |
| 4345.7959 | 4345.7979 | 4345.7999 | 18.665 | 0.029 | 17.945 | 0.023 | 17.035 | 0.021 | 11.795 | 11.075 | 10.165 | 0.351 |
| 4346.5422 | 4346.5442 | 4346.5462 | 18.761 | 0.046 | 17.999 | 0.019 | 17.062 | 0.023 | 11.891 | 11.129 | 10.192 | 0.498 |
| 4346.8033 | 4346.8053 | 4346.8074 | 18.783 | 0.035 | 18.059 | 0.024 | 17.063 | 0.020 | 11.913 | 11.189 | 10.193 | 0.550 |
| 4347.7125 | 4347.7145 | 4347.7166 | 18.807 | 0.028 | 18.098 | 0.024 | 17.156 | 0.025 | 11.937 | 11.228 | 10.286 | 0.733 |
| 4347.8316 | 4347.8336 | 4347.8356 | 18.733 | 0.030 | 18.030 | 0.024 | 17.134 | 0.027 | 11.863 | 11.160 | 10.264 | 0.757 |
| 4348.7182 | 4348.7202 | 4348.7222 | 18.799 | 0.031 | 18.055 | 0.026 | 17.114 | 0.046 | 11.928 | 11.184 | 10.243 | 0.937 |
| 4348.7817 | 4348.7836 | 4348.7857 | 18.819 | 0.033 | 18.081 | 0.022 | 17.144 | 0.024 | 11.948 | 11.210 | 10.273 | 0.950 |
| 4349.6931 | 4349.6951 | 4349.6971 | 18.660 | 0.019 | 17.916 | 0.015 | 16.945 | 0.014 | 11.788 | 11.044 | 10.073 | 1.134 |
| 4349.7773 | 4349.7793 | 4349.7813 | 18.657 | 0.021 | 17.911 | 0.016 | 16.980 | 0.014 | 11.785 | 11.039 | 10.108 | 1.151 |
| 4350.6422 | ... | 4350.6463 | 18.730 | 0.019 | ... | | 16.995 | 0.015 | 11.858 | ... | 10.123 | 1.327 |
| 4350.8212 | 4350.6442 | 4350.8253 | 18.761 | 0.022 | 17.948 | 0.019 | 17.051 | 0.022 | 11.888 | 11.076 | 10.178 | 1.363 |
| 4351.7564 | 4350.8232 | 4351.7605 | 18.695 | 0.030 | 18.022 | 0.022 | 17.157 | 0.026 | 11.821 | 11.149 | 10.283 | 1.552 |
| 4352.7352 | 4351.7584 | 4352.7392 | 18.765 | 0.021 | 17.986 | 0.029 | 17.040 | 0.017 | 11.890 | 11.112 | 10.165 | 1.749 |
| 4353.7808 | 4352.7372 | 4353.7848 | 18.720 | 0.031 | 18.020 | 0.017 | 17.124 | 0.030 | 11.844 | 11.145 | 10.248 | 1.960 |
| 4354.8257 | 4353.7828 | 4354.8297 | 18.846 | 0.029 | 18.008 | 0.026 | 17.245 | 0.021 | 11.968 | 11.132 | 10.367 | 2.169 |
| ... | 4355.7746 | 4355.7767 | ... | | 18.137 | 0.042 | 16.804 | 0.146 | ... | 11.258 | 9.925 | 2.358 |
| 4356.7302 | 4356.7322 | 4356.7342 | 18.834 | 0.023 | 18.093 | 0.017 | 17.202 | 0.018 | 11.953 | 11.212 | 10.321 | 2.548 |
| 4357.7347 | 4357.7367 | 4357.7388 | 18.815 | 0.022 | 18.017 | 0.018 | 17.149 | 0.016 | 11.932 | 11.134 | 10.266 | 2.746 |
| 4358.7164 | 4358.7184 | 4358.7204 | 18.837 | 0.022 | 18.055 | 0.016 | 17.208 | 0.016 | 11.952 | 11.170 | 10.323 | 2.939 |
| 4360.7697 | ... | 4360.7738 | 18.962 | 0.113 | ... | | 17.123 | 0.052 | 12.073 | ... | 10.234 | 3.338 |
| 4362.6857 | 4362.6877 | 4362.6897 | 19.190 | 0.041 | 18.356 | 0.027 | 17.324 | 0.023 | 12.296 | 11.462 | 10.430 | 3.704 |
| 4364.6733 | 4364.6753 | 4364.6773 | 19.117 | 0.080 | 18.194 | 0.042 | 17.385 | 0.064 | 12.217 | 11.294 | 10.485 | 4.079 |
| 4365.6505 | 4365.6525 | 4365.6545 | 18.983 | 0.042 | 18.103 | 0.032 | 17.219 | 0.124 | 12.080 | 11.200 | 10.316 | 4.260 |
| 4370.7035 | 4370.7055 | 4370.7076 | 18.986 | 0.076 | 18.228 | 0.049 | 17.386 | 0.035 | 12.066 | 11.308 | 10.466 | 5.170 |
| 4371.6597 | 4371.6617 | 4371.6638 | 18.899 | 0.047 | 18.302 | 0.033 | 17.349 | 0.026 | 11.976 | 11.379 | 10.426 | 5.336 |
| 4372.6944 | 4372.6964 | 4372.6985 | 19.010 | 0.039 | 18.244 | 0.024 | 17.325 | 0.019 | 12.083 | 11.317 | 10.398 | 5.514 |

**(i) 51378 (2001 AT33)**

| JD (B) | JD (V) | JD (I) | B | $\sigma_B$ | V | $\sigma_V$ | I | $\sigma_I$ | $B_{red}$ | $V_{red}$ | $I_{red}$ | Phase |
|---|---|---|---|---|---|---|---|---|---|---|---|---|
| 4292.7792 | 4292.7811 | 4292.7831 | 19.846 | 0.052 | 19.137 | 0.038 | 18.145 | 0.035 | 13.029 | 12.320 | 11.328 | 2.839 |
| 4297.8408 | 4297.8430 | 4298.7235 | 19.641 | 0.087 | 18.779 | 0.158 | 18.119 | 0.034 | 12.834 | 11.972 | 11.313 | 1.809 |
| 4298.7195 | 4298.7215 | ... | 19.897 | 0.043 | 19.108 | 0.033 | ... | ... | 13.091 | 12.302 | ... | 1.628 |
| 4299.7540 | 4299.7560 | 4299.7580 | 19.933 | 0.039 | 19.127 | 0.035 | 18.124 | 0.034 | 13.128 | 12.322 | 11.319 | 1.414 |
| 4300.7167 | 4300.7186 | 4300.7207 | 19.989 | 0.038 | 19.175 | 0.030 | 18.210 | 0.031 | 13.185 | 12.371 | 11.406 | 1.215 |
| 4301.7707 | 4301.7726 | 4301.7746 | 19.888 | 0.038 | 19.131 | 0.034 | 18.152 | 0.034 | 13.086 | 12.329 | 11.350 | 0.997 |
| 4302.7282 | 4302.7302 | 4302.7322 | 19.866 | 0.047 | 19.170 | 0.035 | 18.157 | 0.032 | 13.064 | 12.368 | 11.355 | 0.800 |
| 4302.7902 | 4302.7922 | 4302.7942 | 19.803 | 0.044 | 19.156 | 0.034 | 18.217 | 0.032 | 13.002 | 12.355 | 11.416 | 0.787 |
| 4304.7293 | 4304.7313 | 4304.7333 | 19.975 | 0.053 | 19.164 | 0.039 | 18.190 | 0.040 | 13.175 | 12.364 | 11.390 | 0.396 |
| 4304.7588 | 4304.7607 | 4304.7628 | 19.920 | 0.072 | 19.023 | 0.052 | 18.169 | 0.054 | 13.120 | 12.223 | 11.369 | 0.390 |
| 4305.7043 | 4305.7063 | 4305.7083 | 19.799 | 0.095 | 18.894 | 0.068 | 17.986 | 0.051 | 12.999 | 12.094 | 11.186 | 0.220 |
| 4305.7432 | 4305.7452 | 4305.7472 | 19.700 | 0.133 | 19.040 | 0.096 | 18.001 | 0.074 | 12.900 | 12.240 | 11.201 | 0.214 |
| 4307.7136 | 4307.7156 | 4307.7176 | 19.917 | 0.076 | 19.104 | 0.044 | 18.129 | 0.034 | 13.118 | 12.305 | 11.330 | 0.300 |
| 4307.7513 | 4307.7533 | 4307.7554 | 19.924 | 0.068 | 19.149 | 0.043 | 18.144 | 0.035 | 13.125 | 12.350 | 11.345 | 0.307 |
| 4308.6467 | 4308.6486 | 4308.6507 | 19.845 | 0.089 | 18.905 | 0.047 | 18.120 | 0.043 | 13.046 | 12.106 | 11.321 | 0.480 |
| ... | 4308.6900 | 4308.6921 | ... | ... | 19.489 | 0.106 | 18.238 | 0.062 | ... | 12.690 | 11.439 | 0.488 |
| 4309.6919 | 4309.6939 | 4309.6959 | 19.950 | 0.146 | 19.032 | 0.088 | 18.134 | 0.087 | 13.150 | 12.232 | 11.334 | 0.691 |
| 4309.7429 | 4309.7448 | 4309.7468 | 20.022 | 0.139 | 19.031 | 0.135 | 18.191 | 0.071 | 13.222 | 12.231 | 11.391 | 0.702 |
| 4312.7454 | 4312.7473 | 4312.7493 | 20.162 | 0.206 | 18.990 | 0.087 | 18.448 | 0.107 | 13.361 | 12.189 | 11.647 | 1.321 |
| 4315.7138 | 4315.7158 | 4315.7178 | 19.853 | 0.097 | 19.058 | 0.052 | 18.061 | 0.059 | 13.049 | 12.254 | 11.257 | 1.934 |
| 4316.6944 | 4316.6964 | 4316.6984 | 19.753 | 0.050 | 19.004 | 0.035 | 18.091 | 0.031 | 12.948 | 12.199 | 11.286 | 2.135 |
| 4319.7203 | 4319.7222 | ... | 19.791 | 0.040 | 19.040 | 0.030 | ... | ... | 12.981 | 12.230 | ... | 2.751 |
| 4320.7184 | 4320.7204 | 4320.7224 | 19.774 | 0.041 | 19.009 | 0.029 | 18.025 | 0.029 | 12.962 | 12.197 | 11.213 | 2.952 |
| 4322.8091 | 4322.8111 | 4322.8131 | 19.814 | 0.073 | 19.124 | 0.058 | 18.189 | 0.071 | 12.998 | 12.308 | 11.373 | 3.369 |
| 4326.7450 | 4326.7470 | 4326.7490 | 19.816 | 0.051 | 19.100 | 0.036 | 18.128 | 0.037 | 12.990 | 12.274 | 11.302 | 4.137 |
| 4330.7181 | 4330.7200 | 4330.7221 | 19.774 | 0.040 | 19.020 | 0.034 | 18.046 | 0.034 | 12.936 | 12.182 | 11.208 | 4.884 |
| 4332.7624 | 4332.7644 | 4332.7664 | 19.831 | 0.064 | 19.051 | 0.046 | 18.117 | 0.045 | 12.986 | 12.206 | 11.272 | 5.256 |
| 4334.7178 | 4334.7198 | 4334.7218 | 19.681 | 0.058 | 19.077 | 0.044 | 18.096 | 0.038 | 12.829 | 12.225 | 11.244 | 5.604 |
| 4335.6537 | 4335.6557 | 4336.6241 | 19.752 | 0.085 | 18.991 | 0.057 | 18.024 | 0.056 | 12.897 | 12.135 | 11.165 | 5.767 |
| 4336.6200 | 4336.6220 | 4340.6927 | 19.649 | 0.143 | 18.963 | 0.095 | 18.044 | 0.059 | 12.790 | 12.104 | 11.167 | 5.933 |
| 4340.6887 | 4340.6907 | 4341.7024 | 19.687 | 0.135 | 18.973 | 0.084 | 18.209 | 0.053 | 12.810 | 12.096 | 11.328 | 6.608 |
| 4341.6984 | 4341.7004 | 4347.7067 | 19.675 | 0.126 | 19.145 | 0.117 | 18.293 | 0.097 | 12.794 | 12.264 | 11.383 | 6.769 |
| 4347.7027 | 4347.7047 | 4352.6665 | 19.957 | 0.101 | 19.106 | 0.082 | 18.230 | 0.048 | 13.047 | 12.196 | 11.293 | 7.668 |
| 4352.6625 | 4352.6645 | 4356.6813 | 19.942 | 0.047 | 19.173 | 0.035 | 18.272 | 0.046 | 13.005 | 12.236 | 11.313 | 8.332 |
| 4356.6773 | 4356.6793 | 4361.6563 | 19.935 | 0.054 | 19.159 | 0.044 | 17.875 | 0.040 | 12.976 | 12.200 | 10.886 | 8.815 |
| 4361.6523 | 4361.6543 | 4370.6416 | 19.980 | 0.087 | 19.043 | 0.050 | 18.655 | 0.092 | 12.991 | 12.054 | 11.610 | 9.341 |
| 4371.6016 | 4370.6395 | 4371.6057 | 20.331 | 0.152 | 19.688 | 0.139 | 18.710 | 0.065 | 13.280 | 12.643 | 11.659 | 10.091 |
| 4373.5823 | 4371.6036 | 4373.5864 | 20.465 | 0.063 | 19.693 | 0.096 | 18.828 | 0.055 | 13.401 | 12.642 | 11.764 | 10.155 |
| 4375.5706 | 4373.5843 | 4375.5747 | 20.386 | 0.138 | 19.710 | 0.045 | 18.881 | 0.098 | 13.309 | 12.646 | 11.804 | 10.280 |

## Table 3. Summary of Phase Curve Fits

| | $N_{obs}$ | $\sigma_{sys}$ | Linear Fit | | | | | Hapke Fit | | | |
|---|---|---|---|---|---|---|---|---|---|---|---|
| | | | S | $m_0$ | $\chi 2$ | $S_{\leq 2}$ | $S_{>5}$ | $m_0$ | $h_c$ | $B_{C0}$ | $\chi 2$ |
| 588 Achilles | 38 | 0.02 | 0.04±0.01 | 8.41 | 36.5 | 0.05±0.02 | 0.04±0.02 | 8.40±0.01 | 10.3±0.5 | 1 | 44.9 |
| 1208 Troilus | 12 | 0.02 | 0.05±0.02 | 9.08 | 10.1 | 0.06±0.03 | … | 9.08±0.01 | 7.4±0.5 | 1 | 10.5 |
| 4348 Poulydamas | 9 | 0.08 | 0.09±0.04 | 9.75 | 7.1 | 0.04±0.09 | … | 9.74±0.03 | 4.7±1.0 | 1 | 11.8 |
| 6998 Tithonus | 11 | 0.06 | 0.06±0.02 | 11.66 | 9.4 | 0.09±0.07 | … | 11.65±0.02 | 8.0±1.5 | 1 | 9.0 |
| 8317 Eurysaces | 104 | 0.09 | 0.06±0.01 | 11.33 | 102.2 | 0.00±0.02 | 0.04±0.01 | 11.27±0.01 | 4.5±0.4 | 1 | 105.9 |
| 12126 (1999 RM11) | 14 | 0.09 | 0.05±0.02 | 10.36 | 12.1 | 0.05±0.11 | … | 10.36±0.02 | 8.7±1.5 | 1 | 13.3 |
| 13323 (1998 SQ) | 54 | 0.09 | 0.04±0.01 | 11.30 | 54.1 | 0.08±0.04 | 0.04±0.01 | 11.27±0.01 | 7.8±1.0 | 1 | 52.7 |
| 24506 (2001 BS15) | 41 | 0.08 | 0.05±0.01 | 11.00 | 39.7 | 0.00±0.02 | 0.07±0.02 | 11.00±0.01 | 8.4±1.1 | 1 | 43.8 |
| 51378 (2001 AT33) | 35 | 0.12 | -0.02±0.01 | 12.23 | 33.1 | -0.01±0.06 | 0.00±0.01 | … | … | … | … |

**Table 4. Comparison of Trojan Surface Properties with Other Bodies**

| | $S_{\leq 2}$ (mag/°) | $N_{\leq 2}$ | ref | $S_{>5}$ (mag/°) | $N_{>5}$ | ref | B-R (mag) | ref | albedo (%) | ref |
|---|---|---|---|---|---|---|---|---|---|---|
| **Small/Gray Bodies** | | | | | | | | | | |
| Trojans | 0.00-0.09 | 8 | 1 | 0.04-0.07 | 5 | 1 | 1.0-1.5 | 4 | 3-6 | 8 |
| Gray Centaurs | 0.01-0.17 | 8 | 2 | … | | | 1.0-1.4 | 5 | 3-6 | 9 |
| Gray SDOs | 0.18 | 1 | 2 | … | | | 1.1-1.6 | 5 | 3-7 | 9 |
| Dead Comet Candidates | … | | | … | | | 1.0-1.5 | 6 | 2-9 | 8 |
| **Transition Bodies** | | | | | | | | | | |
| Active Centaurs | 0.18 | 1 | 2 | … | | | 1.0-1.4 | 5 | 4-7 | 9 |
| Active Comet Nuclei | … | | | … | | | 0.9-1.6 | 7,12 | 2-6 | 7 |
| **Small/Red Bodies** | | | | | | | | | | |
| Red Centaurs | 0.18 | 1 | 2 | … | | | 1.7-2.0 | 7 | 7-18 | 9 |
| KBOs | 0.10-0.25 | 13 | 2 | … | | | 1.5-2.0 | 7 | 6-16 | 9 |
| **Main Belt Asteroids** | | | | | | | | | | |
| D Class (non-Trojan) | … | | | … | | | 1.4-1.5 | 11 | 4-6 | 11 |
| P Class | 0.075-0.113 | 4 | 3 | 0.039-0.044 | 4 | 3 | 1.2-1.5 | 11 | 2-7 | 11 |
| C Class | 0.065-0.127 | 9 | 3 | 0.039-0.046 | 9 | 3 | 1.2-1.6 | 10 | 3-9 | 11 |
| M Class | 0.145-0.197 | 6 | 3 | 0.029-0.036 | 6 | 3 | 1.3-1.5 | 10 | 7-30 | 11 |
| S Class | 0.151-0.196 | 9 | 3 | 0.024-0.032 | 9 | 3 | 1.5-2.0 | 10 | 4-50 | 11 |
| E Class | 0.084-0.138 | 4 | 3 | 0.018-0.022 | 4 | 3 | 1.1-1.5 | 10 | 16-55 | 11 |

References: (1) This work, not including 51378 (2001 AT33), which had an anomalous slope; (2) Schaefer et al. (2009a); (3) derived from individual fit parameters for each asteroid as reported in Belskaya and Shevchenko (2000); (4) Fornasier et al. (2007); (5) Tegler et al. (2003); (6) Jewitt (2002); (7) Lamy et al. (2004); (8) Fernandez et al. (2003); (9) Stansberry et al. (2008); (10) B-R values are calculated from the R/B values in Zellner (1979) using the formula B-R = 2.5 log (R/B) + (B-R)$_{sun}$, with a value for (B-R)$_{sun}$ of 1.19; (11) Compiled from the JPL Small-Body Database (http://ssd.jpl.nasa.gov/sbdb_query.cgi#x), with R/B values taken from Chapman and Gaffey (1979); B-R values were calculated as in (10); (12) Lamy and Toth (2009).

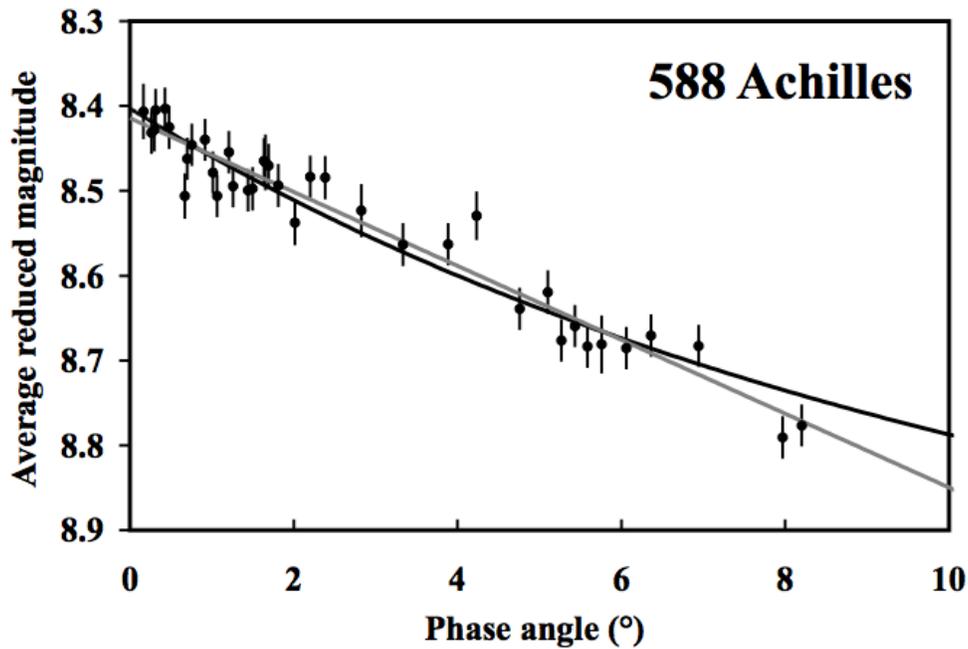

**Figure 1.** This phase curve for 588 Achilles is representative of all nine Trojan phase curves. It is a plot of the average reduced magnitude ($m_{avg}$) versus the solar phase angle ($\alpha$). The gray line represents the best linear fit, and the black line represents the best Hapke fit model. For Achilles, we have 38 magnitudes that consist of averages of our nearly-simultaneous B-, V-, and I-band magnitudes with corrections to a solar distance and geocentric distance of 1 AU. The one-sigma error bars plotted are the combined uncertainty from the measurement errors added in quadrature with an arbitrary systematic error ($\sigma_{sys}$, 0.02 mag in this case).

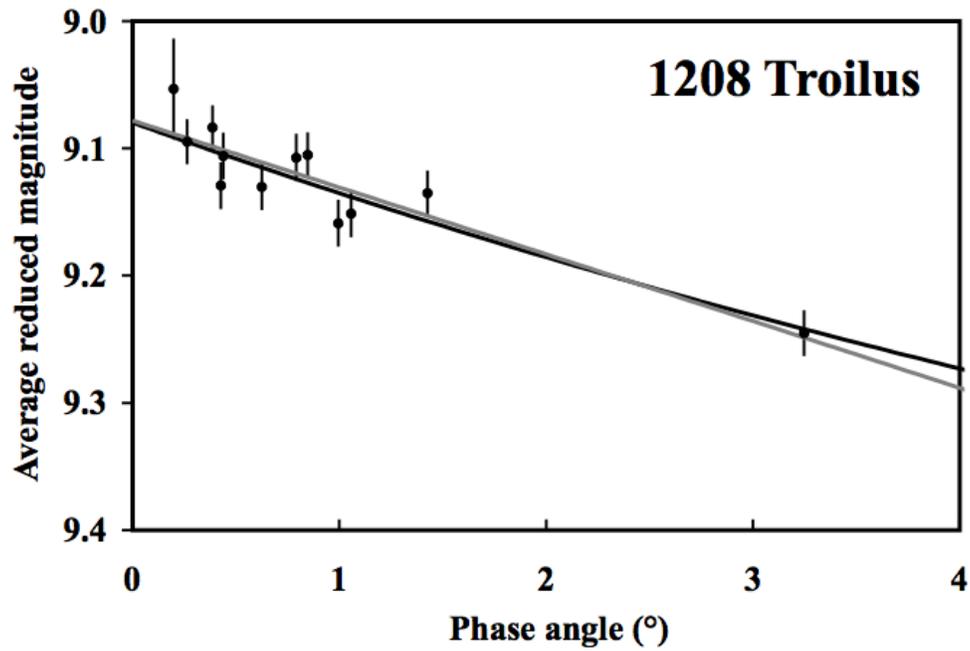

**Figure 2.** This phase curve for 1208 Troilus is the poorest-sampled in phase of our program. Nevertheless, the slope is well-measured, with a one-sigma uncertainty of ±0.02 mag/°. This slope is far larger than is possible by shadow-hiding alone, proving that coherent backscattering dominates in the opposition surge for this body.

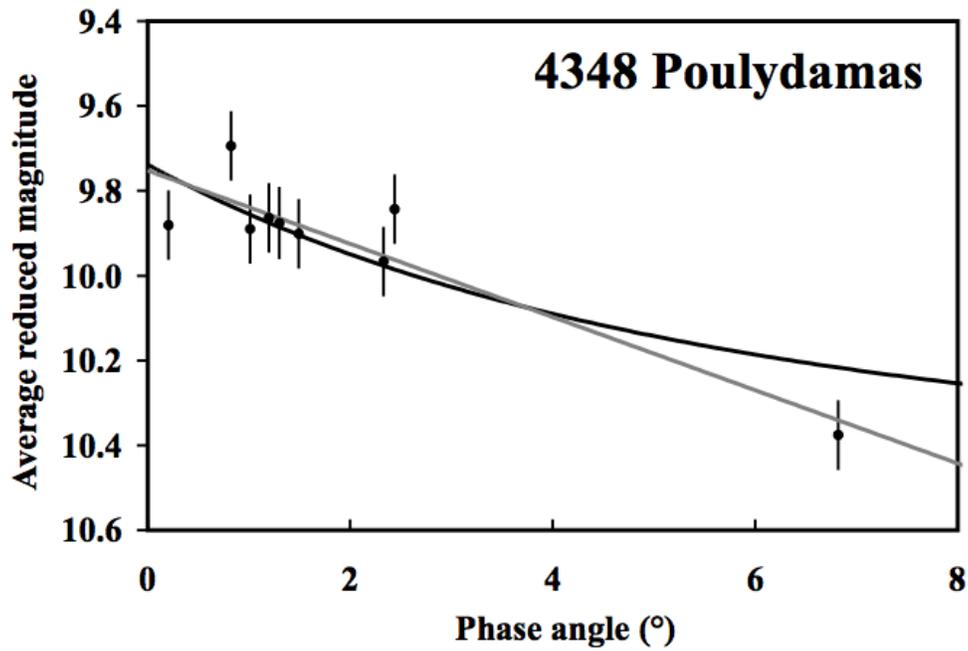

**Figure 3.** This phase curve for 4348 Poulydamas is easily consistent with a simple empirical linear fit. The linear model fits well ($\chi^2$=7.1) and the Hapke model fits less well ($\chi^2$=11.8). This difference in $\chi^2$ shows that the linear fit is preferred at the 2-sigma confidence level. Almost all of this difference is caused by the one point at high phase angle where the Hapke model just can't get the value low enough due to the physical constraint that $B_{C0} \leq 1$.

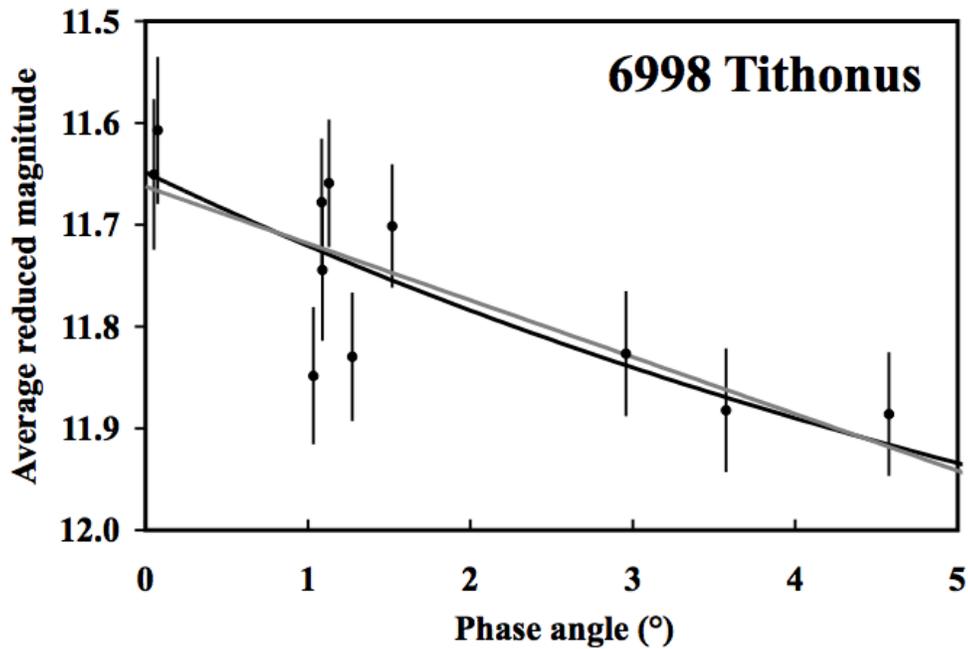

**Figure 4.** The phase curve of 6998 Tithonus appears to have a lot of scatter. With the given error bars, we see that nine out of eleven points are in agreement with either model to within one sigma. But this is illusory, because the quoted error bars have had an arbitrary uncertainty of $\sigma_{sys}$=0.06 mag added in quadrature to the evaluated measurement errors. Therefore, apparently there is an extra variance in our data for Tithonus that has a typical scatter of 0.06 mag. We think that the most likely cause for this extra variance is uncorrected rotational modulation.

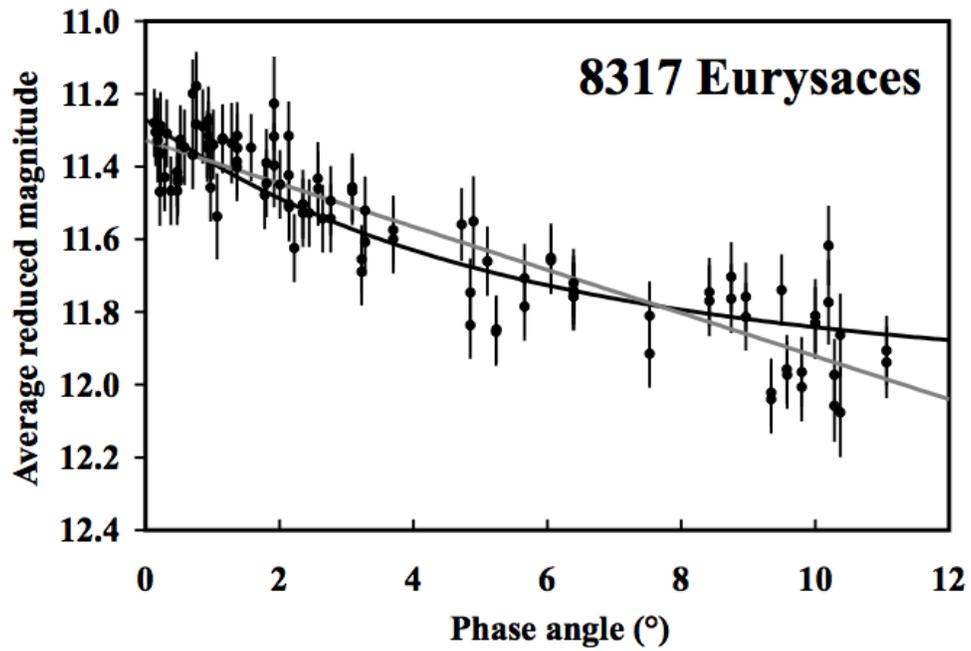

**Figure 5.** This phase curve for 8317 Eurysaces is constructed from two observing seasons (in 2006 and 2007) including 228 individual magnitude measurements (in B, V, and I) averaged together to for 104 points. This is our best-sampled phase curve. Even so, we could not pull out a rotational period with any confidence, primarily because our observing cadence was optimized for phase curves rather than rotational curves.

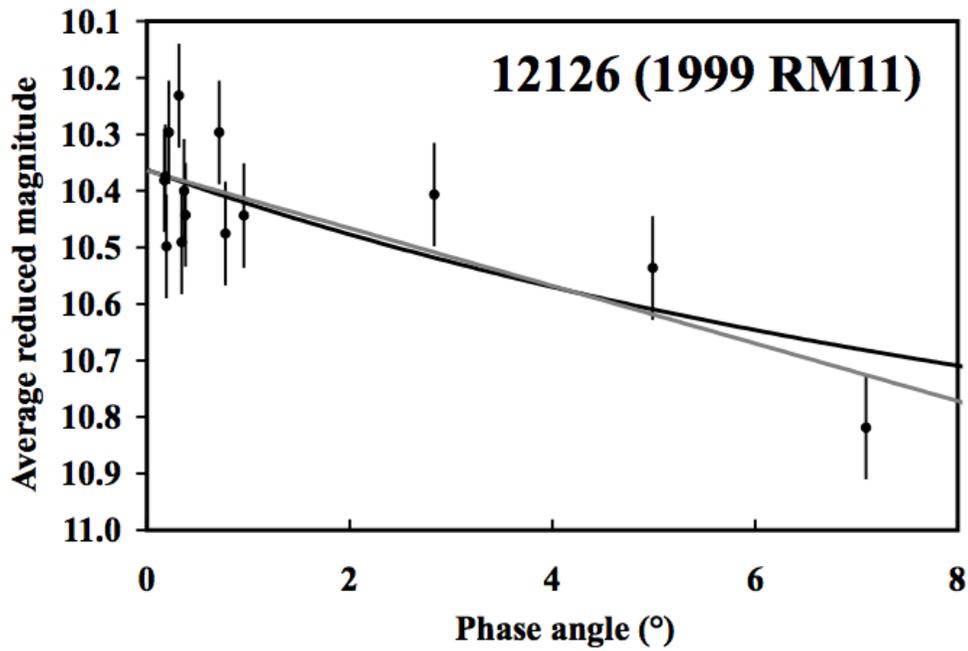

**Figure 6.** This phase curve of 12126 (1999 RM11) shows no significant difference between the linear and Hapke models. Indeed, all of our phase curves have no significant difference (at better than a 3-sigma confidence level) in $\chi^2$ between the two models. This is to say that all our phase curves are adequately described by a simple straight line with no apparent curvature.

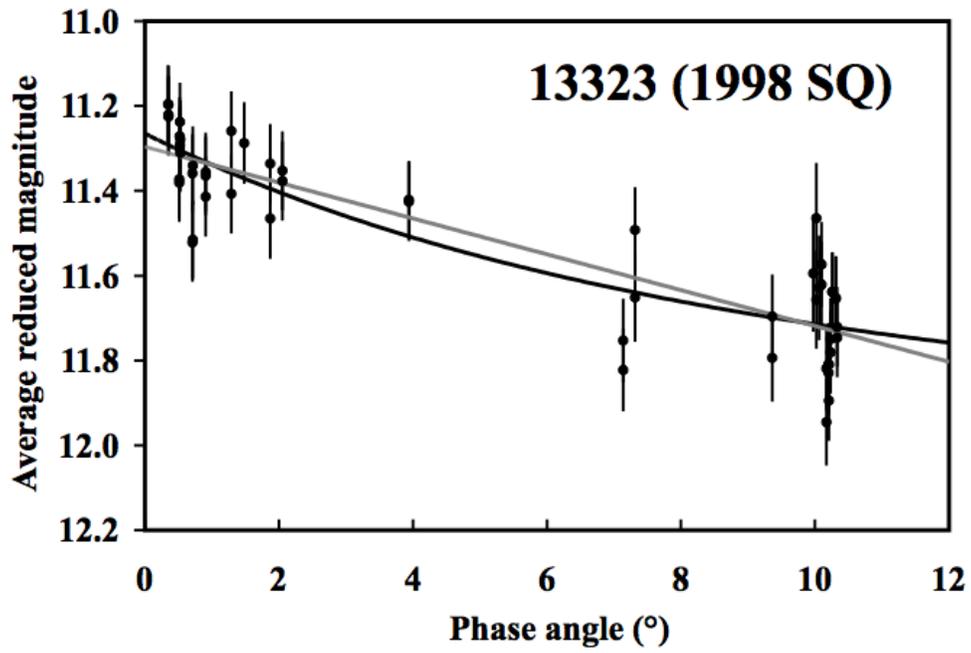

**Figure 7.** Again, in the phase curve for 13323 (1998 SQ) we see no significant difference here between the linear fit and the Hapke fit.

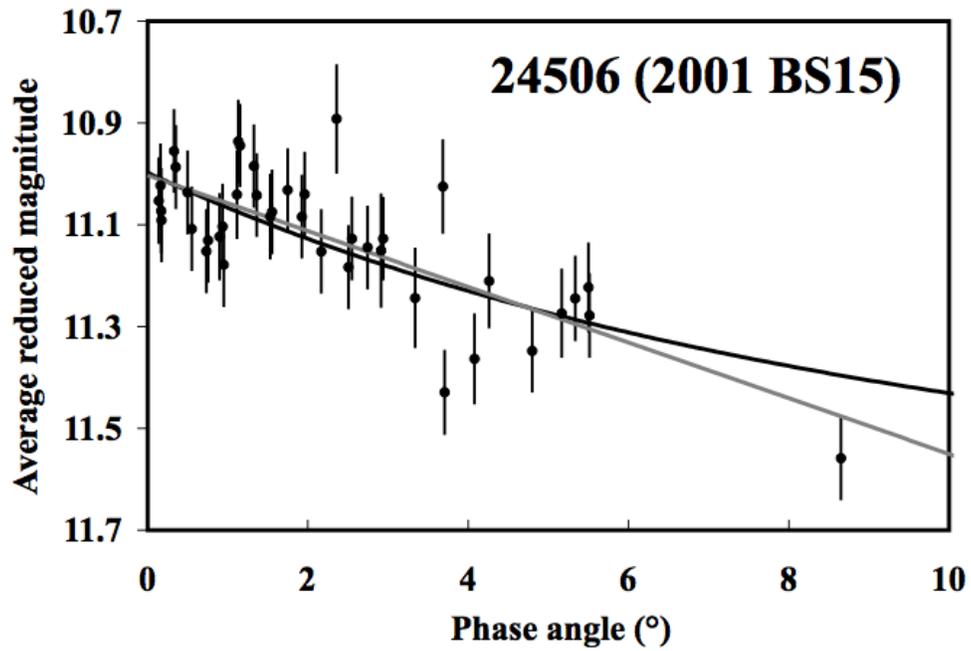

**Figure 8.** This phase curve for 24506 (2001 BS15) shows a small preference for the linear fit ($\chi^2$=39.7) over the Hapke model ($\chi^2$=43.8). Just as for 4348 Poulydamas (see Figure 3), the difference in $\chi^2$ arises almost solely from the one high-phase point. Again, the reason why the Hapke model cannot better fit that point is the physical constraint that $B_{C0} \leq 1$.

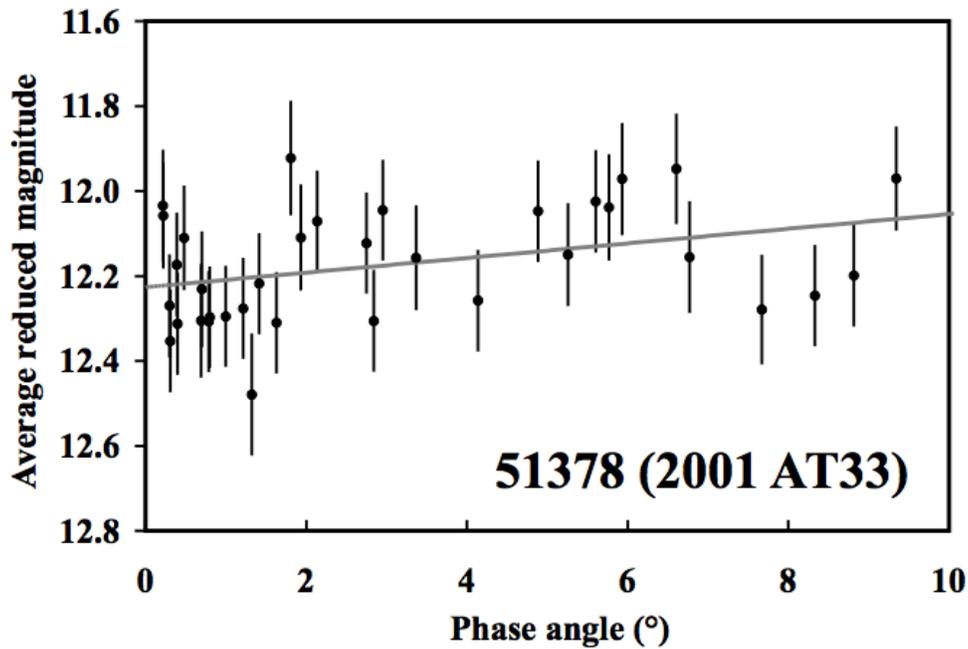

**Figure 9.** This phase curve for 51378 (2001 AT33) formally returns a negative slope from a linear fit. We know of no precedent or physical mechanism that can yield a negative slope. Likely, this Trojan simply has a near-zero opposition surge and ordinary variations (perhaps associated with rotational modulation) randomly result in a marginally negative slope. This idea is supported by the relatively large scatter of individual points (with $\sigma_{sys}$=0.12 mag) that would arise for common rotational amplitudes if uncorrected.